\documentclass{SciPost}
\usepackage{bm}
\hypersetup{
    colorlinks,
    linkcolor={red!50!black},
    citecolor={blue!50!black},
    urlcolor={blue!80!black}
}

\usepackage[bitstream-charter]{mathdesign}
\DeclareSymbolFont{usualmathcal}{OMS}{cmsy}{m}{n}
\DeclareSymbolFontAlphabet{\mathcal}{usualmathcal}

\fancypagestyle{SPstyle}{
\fancyhf{}
\lhead{\colorbox{scipostblue}{\bf \color{white} ~SciPost Physics Codebases }}
\rhead{{\bf \color{scipostdeepblue} ~Submission }}

\fancyfoot[C]{\textbf{\thepage}}
}

\newcommand{\zmin}{z_{2,\textrm{min}}}

\newcommand{\sigmaltnlo}{\sigma_{L,T}^{\textrm{NLO}}}
\newcommand{\sigmaltdip}{\sigma_{L,T}^{\textrm{dip}}}
\newcommand{\sigmaltqg}{\sigma_{L,T}^{q\bar q g}}

\newcommand{\sigmaltqgu}{\sigma_{L,T}^{q\bar qg}}
\newcommand{\sigmalt}[1]{\sigma_{L,T}^{#1}}
\newcommand{\aem}{\alpha_\mathrm{em}}
\newcommand{\as}{\alpha_\mathrm{s}}
\newcommand{\cf}{C_\mathrm{f}}
\newcommand{\nc}{{N_\mathrm{c}}}
\newcommand{\nf}{{N_\mathrm{F}}}
\newcommand{\xt}{{\mathbf{x}}}
\newcommand{\bt}{{\mathbf{b}}}
\newcommand{\Tr}{\mathrm{Tr}}

\newcommand{\kaz}{\kappa_{z}}
\newcommand{\kac}{\kappa_{\chi}}
\newcommand{\hcal}{\mathcal{H}}
\newcommand{\xbj}{x_\mathrm{Bj}}

\usepackage[caption=false]{subfig}

\usepackage{tikz}
\usepackage{physics}

\usepackage{listings}
\usepackage{xcolor}

\lstdefinestyle{inlinecpp}{
    language=C++,
    basicstyle=\ttfamily\small,
    keywordstyle=\color{blue},
    commentstyle=\color{gray},
    stringstyle=\color{red},
    showstringspaces=false,
    breaklines=true,
}

\lstdefinestyle{bashstyle}{
    backgroundcolor=\color{gray!10},
    basicstyle=\ttfamily\small,
    frame=single,
    breaklines=true,
    columns=fullflexible,
    showstringspaces=false
}

\newcommand{\code}[1]{\lstinline[style=inlinecpp]!#1!}
\lstnewenvironment{cli}{
    \lstset{style=bashstyle}
}{}

\lstdefinestyle{cppstyle}{
    language=C++,
    backgroundcolor=\color{gray!8},
    basicstyle=\ttfamily\small,
    keywordstyle=\color{blue},
    commentstyle=\color{green!50!black},
    stringstyle=\color{red!60!black},
    numbers=left,
    numberstyle=\tiny\color{gray},
    stepnumber=1,
    numbersep=8pt,
    frame=single,
    breaklines=true,
    tabsize=4,
    showstringspaces=false
}

\lstnewenvironment{cppcode}{
    \lstset{style=cppstyle}
}{}

\begin{document}

\pagestyle{SPstyle}

\begin{center}{\Large \textbf{\color{scipostdeepblue}{
A numerical implementation of the NLO DIS structure functions in the dipole picture\\
}}}\end{center}

\begin{center}\textbf{
Henri Hänninen\textsuperscript{$1^\dagger$}
Heikki Mäntysaari\textsuperscript{$2,3^\star$}
Jani Penttala\textsuperscript{$4,5^\ddagger$}
}\end{center}

\begin{center}
{\bf 1} Department of Mathematics and Statistics, University of Jyväskylä,
 P.O. Box 35, 40014 University of Jyv\"askyl\"a, Finland
\\
{\bf 2} Department of Physics, University of Jyväskylä,  P.O. Box 35, 40014 University of Jyväskylä, Finland
\\
{\bf 3} Helsinki Institute of Physics, P.O. Box 64, 00014 University of Helsinki, Finland
\\
{\bf 4} Department of Physics and Astronomy, University of California, Los Angeles, CA 90095, USA
\\
{\bf 5} Mani L. Bhaumik Institute for Theoretical Physics, University of California, Los Angeles, CA 90095, USA
\\[\baselineskip]
$\star$ \href{mailto:heikki.mantysaari@jyu.fi}{\small heikki.mantysaari@jyu.fi}\,,\quad
$\dagger$ \href{mailto:henri.j.hanninen@jyu.fi}{\small henri.j.hanninen@jyu.fi}\,,\quad
$\ddagger$ \href{mailto:janipenttala@physics.ucla.edu}{\small janipenttala@physics.ucla.edu}
\end{center}

\section*{\color{scipostdeepblue}{Abstract}}
\textbf{\boldmath{%
We present a numerical program that evaluates deep inelastic scattering (DIS) structure functions at next-to-leading order (NLO) accuracy in the dipole picture. In this numerical implementation the NLO DIS impact factors with massive quarks are written in a form that ensures a stable numerical evaluation of the DIS cross sections.  
}
}

\vspace{\baselineskip}

\noindent\textcolor{white!90!black}{%
\fbox{\parbox{0.975\linewidth}{%
\textcolor{white!40!black}{\begin{tabular}{lr}%
  \begin{minipage}{0.6\textwidth}%
    {\small Copyright attribution to authors. \newline
    This work is a submission to SciPost Physics Codebases. \newline
    License information to appear upon publication. \newline
    Publication information to appear upon publication.}
  \end{minipage} & \begin{minipage}{0.4\textwidth}
    {\small Received Date \newline Accepted Date \newline Published Date}%
  \end{minipage}
\end{tabular}}
}}
}


\vspace{10pt}
\noindent\rule{\textwidth}{1pt}
\tableofcontents
\noindent\rule{\textwidth}{1pt}
\vspace{10pt}

\section{Introduction}
\label{sec:intro}

Deep inelastic scattering (DIS), in which an electron scatters off a proton by exchanging a virtual photon (or a $Z$ boson at high virtualities) provides a clean environment to study the partonic structure of hadrons. Indeed, precision measurements performed at the DESY-HERA $e+p$-collider have enabled precise extractions of the proton quark and gluon distributions~\cite{H1:2015ubc}. One striking feature revealed by HERA measurements is the rapid rise of the gluon density towards the small momentum fraction $x$, a rise that cannot continue indefinitely without violating unitarity.

At small Bjorken-$\xbj$, accessible in high-energy scattering experiments, saturation phenomena described in the Color Glass Condensate (CGC) effective theory of high-energy QCD~\cite{Iancu:2003xm,Garcia-Montero:2025hys} provide a natural mechanism to tame this growth. In the frame where the proton longitudinal momentum is very large, these phenomena can be interpreted as gluon fusion $g+g\to g$, a non-linear process slowing down the growth of the proton density at small-$\xbj$. Although saturation effects are genuinely predicted by QCD, and several hints have been observed in various scattering experiments in DIS and proton-nucleus collisions, no conclusive evidence of gluon saturation has been obtained to date, see Ref.~\cite{Morreale:2021pnn} for a review. 

In order to find conclusive evidence of saturation effects, it is crucial to match the precision of the theoretical calculations to that of the experimental data. The HERA measurements of the total DIS cross section mentioned above are very precise, and in the next decade similar precision measurements of electron-nucleus scattering will be performed at the Electron-Ion Collider (EIC)~\cite{AbdulKhalek:2021gbh}.

In this work we develop a general purpose numerical program that, when given as input the dipole-target scattering amplitude evolved by the Balitsky-Kovchegov equation at next-to-leading order~\cite{Lappi:2016fmu,Balitsky:2008zza,Iancu:2015vea,Iancu:2015joa}, can be used to evaluate the DIS structure functions in the dipole picture~\cite{Mueller:1994jq} at next-to-leading order accuracy. 
Specifically, we implement the NLO DIS impact factors with massive quarks calculated in Refs.~\cite{Beuf:2022ndu,Beuf:2021qqa,Beuf:2021srj}. Similar results in the massless quark limit are also available~\cite{Balitsky:2010ze, Beuf:2011xd, Balitsky:2012bs,Beuf:2016wdz, Hanninen:2017ddy, Beuf:2017bpd}; however explicit results in the $m_q=0$ limit have not been included in this program. Although numerical calculations are easier in the zero mass limit, we have confirmed that the developed program matches smoothly the zero quark mass results when the quark mass is set to be small. In most phenomenological applications retaining a finite quark mass is also advantageous because it  suppresses the aligned-jet contributions.

\section{Deep inelastic scattering at next-to-leading order in dipole picture}

The total cross section for the electron-proton (or electron-nucleus) deep inelastic scattering is typically expressed in terms of the reduced cross section $\sigma_r$ defined as
\begin{equation}
    \sigma_r(y,\xbj,Q^2) = F_2\left(\xbj,Q^2\right) - \frac{y^2}{1+(1-y)^2} F_L\left(\xbj,Q^2\right).
\end{equation}
Here $y=Q^2/(s\xbj)$ is the inelasticity variable, $F_2=F_T+F_L$, and the structure functions $F_{2,T,L}$ are proportional to the total cross section for the virtual photon-proton scattering:
\begin{align}
    F_2 &= \frac{Q^2}{4\pi^2 \aem} (\sigma^{\gamma^* p}_T + \sigma^{\gamma^* p}_L), \\
    F_{T,L} &= \frac{Q^2}{4\pi^2 \aem} \sigma^{\gamma^* p}_{T,L}.
\label{eq:f2fl}
\end{align}
Here $T$ and $L$ denote the transverse and longitudinal virtual photon polarization states, respectively, $Q^2=-q^2$ where $q$ is the photon four momentum, and $\aem$ is the fine-structure constant. 

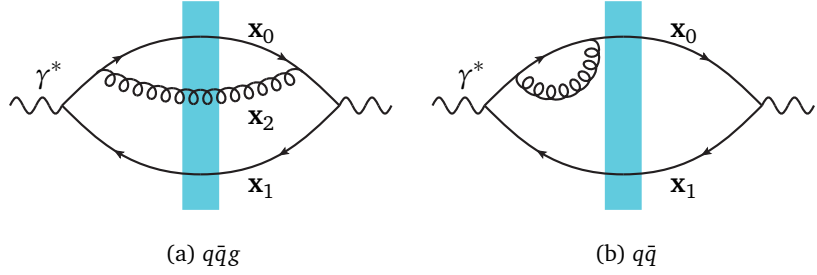
\begin{figure}
     \centering
    \subfloat[$q\bar q g$\label{fig:qqg}]{
         \includegraphics[width=0.35\columnwidth]{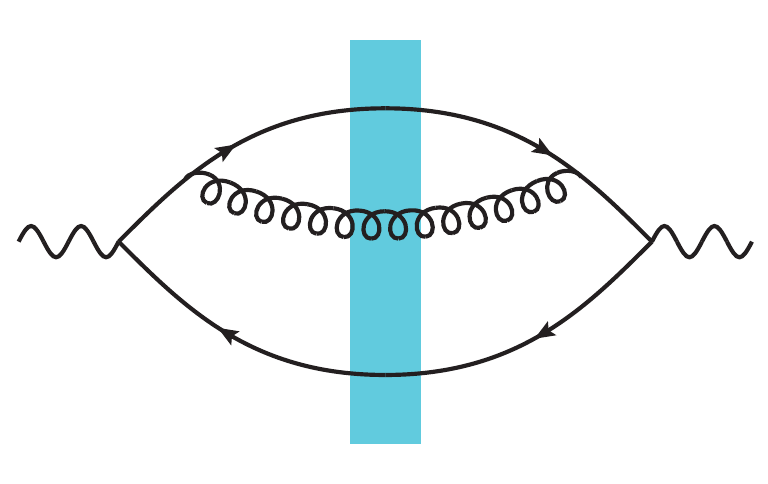}
          \begin{tikzpicture}[overlay]
         \node[anchor=south east] at (-1.6cm,2.35cm) {$\xt_0$};
         \node[anchor=south east] at (-1.6cm,0.3cm) {$\xt_1$};
         \node[anchor=south east] at (-1.6cm,1.2cm) {$\xt_2$};
        \node[anchor=south east] at (-4.4cm,1.75cm) {$\gamma^*$};
        \end{tikzpicture}
         }
    \subfloat[$q\bar q$\label{fig:qq}]{
         \includegraphics[width=0.35\columnwidth]{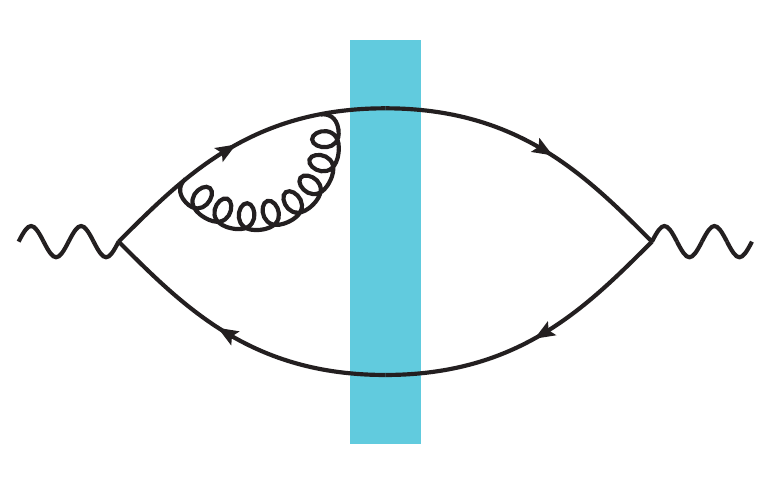}
         \begin{tikzpicture}[overlay]
         \node[anchor=south east] at (-1.6cm,2.35cm) {$\xt_0$};
         \node[anchor=south east] at (-1.6cm,0.3cm) {$\xt_1$};
         \node[anchor=south east] at (-4.4cm,1.75cm) {$\gamma^*$};
         \end{tikzpicture}
    }
     \caption{
    Example diagrams contributing to the elastic $\gamma^*+p \to \gamma^* + p$ amplitude at NLO. The blue band represents the interaction with the target shockwave. Figure from Ref.~\cite{Hanninen:2022gje}.
     }
     \label{fig:diags}
\end{figure}
The total DIS cross section is typically calculated using the optical theorem, evaluating the $\gamma^*p \to \gamma^* p$ forward elastic scattering amplitude (see however Ref.~\cite{Bertilsson:2026vtu}). In the NLO calculations this is typically divided into different parts depending on the partonic state interacting with the target shockwave. In the unsubtracted scheme developed in Ref.~\cite{Ducloue:2017ftk} that we follow in this work, this corresponds to the separation 
\begin{equation}
\label{eq:nloxs}
    \sigmaltnlo = \sigmalt{\textrm{IC}} + \sigmaltdip + \sigmaltqgu .
\end{equation}
From now on we drop the superscript $\gamma^*p$ for brevity.
The different contributions correspond to  convolutions of hard factors describing the perturbative splitting $\gamma^* \to q\bar q$ or $\gamma^* \to q\bar q g$ (and similarly the reverse processes), and Wilson lines $V(\xt)$ ($U(\xt)$) describing an eikonal propagation of quarks (gluons) at fixed transverse coordinate $\xt$ in the target color field. 

The Wilson line correlators appearing in the NLO DIS cross section satisfy the perturbative Balitsky-Kovchegov (BK)~\cite{Kovchegov:1999yj,Balitsky:1995ub} or JIMWLK~\cite{Jalilian-Marian:1997qno,Iancu:2000hn,Mueller:2001uk,Lappi:2012vw,Cali:2021tsh} equation describing their energy (or $\xbj$) dependence. The BK equation is currently known at NLO accuracy~\cite{Balitsky:2008zza}, with the most important higher order corrections enhanced by large transverse logarithms resummed to all orders in Refs.~\cite{Iancu:2015vea,Iancu:2015joa,Beuf:2014uia}. The non-perturbative initial condition for this evolution must be determined by fitting experimental data, typically the total DIS cross section from HERA, as done in Refs.~\cite{Beuf:2020dxl,Hanninen:2022gje,Casuga:2025etc}. The numerical program developed in this work takes the dipole-target scattering amplitude as external input, and can directly use datafiles published as part of these existing fits.

In Eq.~\eqref{eq:nloxs}, the first term corresponds to the lowest-order contribution where only the $|q\bar q\rangle$ state of the virtual photon is included, and no BK/JIMWLK evolution is involved in the description of the target, i.e. Wilson line correlators are evaluated at the initial condition of the small-$\xbj$ evolution. The second contribution, $\sigmaltdip$, corresponds to the loop corrections to the process where a $q\bar q$ system interacts with the shockwave. For the $q\bar q$-target  scattering, the amplitude is given by the correlator of two Wilson lines
\begin{equation}
    N_{01} = 1-\frac{1}{\nc} \left\langle  \Tr{V(\xt_0) V^\dagger(\xt_1)} \right\rangle . \label{eq:s01}
\end{equation}
Here $\xt_0$ and $\xt_1$ are the quark and antiquark coordinates and $\langle \mathcal{O}\rangle$ refer to the average over target configurations. 
Similarly, the case where a $q\bar qg $ system interacts with the target is included in the term $\sigmaltqg$, in which case the interaction with the shockwave reads
\begin{equation}
     N_{012} = 1-\frac{\nc}{2\cf} \left( S_{02}S_{21} - \frac{1}{\nc^2} S_{01}\right), \label{eq:s012}
\end{equation}
where $S_{ij} = 1-N_{ij}$.
Example diagrams for both classes of contributions are shown in Fig.~\ref{fig:diags}. 

As already mentioned above, the Wilson lines satisfy the BK/JIMWLK evolution equation and as such depend implicitly on the evolution rapidity $Y$. Following Refs.~\cite{Beuf:2020dxl,Casuga:2025etc}, the Wilson line correlators in $\sigmalt{\textrm{IC}}$ are evaluated at $Y=Y_0$ where $Y_0$ is the rapidity at which the initial condition is parametrized. In  $\sigmaltdip$ the evolution rapidity is 
\begin{equation}
    Y_\mathrm{dip} = \ln \frac{1}{x}.
\end{equation}
There are two possible choices for the momentum fraction $x$. First, one can use 
\begin{equation}
    x=x_\mathrm{Bj}=\frac{Q^2}{Q^2+W^2}
\end{equation}
by setting the heavy quark momentum fraction scheme as
\begin{cppcode}
NLODIS::SetHeavyQuarkXScheme(HeavyQuarkX::MassIndependentX)
\end{cppcode}
Here \code{NLODIS} is the C++ class that implements the NLO cross section calculation, see Sec.~\ref{sec:usage}.
This is the choice made e.g. in NLO fits Refs.~\cite{Casuga:2025etc,Hanninen:2022gje}. The other option is to take into account the finite quark mass as e.g. in Ref.~\cite{Albacete:2010sy}, in which case
\begin{equation}
    x = \frac{Q^2+4m^2}{Q^2+W^2}.
\end{equation}
In option can be enabled by calling
\begin{cppcode}
NLODIS::SetHeavyQuarkXScheme(HeavyQuarkX::MassDependentX)
\end{cppcode}
When evaluating the $\sigmaltqgu$ term the evolution rapidity is~\cite{Ducloue:2017ftk,Beuf:2020dxl,Casuga:2025etc}
\begin{equation}
    Y=\ln \frac{W^2 z_2}{Q_0^2}.
\end{equation}
Here $z_2$ is the fraction of the photon large plus momentum carried by the gluon, see Sec.~\ref{sec:explicit_xs}.
Here $Q_0^2$ is a characteristic non-perturbative transverse momentum scale of the target, that is typically set to $Q_0^2=1\,\mathrm{GeV}^2$. In the code this constant is \code{NLODIS::config.Q0sqr}.
The evolution rapidities are evaluated in 
\begin{cppcode}
NLODIS::EvolutionRapidity_qqg(double xbj, double Q2, double z2) const;
NLODIS::EvolutionRapidity_dipole(double xbj, double Q2, double quark_mass) const;
\end{cppcode}

Explicit expressions for the contributions  $\sigmalt{\textrm{IC}},  \sigmaltdip$ and $ \sigmaltqgu$ at NLO have have been obtained in Refs.~\cite{Beuf:2022ndu,Beuf:2021qqa,Beuf:2021srj}. For completeness, these results are repeated in Sec.~\ref{sec:explicit_xs}, where we also present them in a form that is more suitable for numerical implementation. We note that although the NLO DIS cross section is independent of the scheme used to cancel the ultraviolet divergence between the two NLO terms, the individual $\sigmaltqgu$ and $\sigmaltdip$ contributions are UV-subtraction-scheme dependent. In this work, we use the exponential subtraction scheme from Ref.~\cite{Hanninen:2017ddy} as in Refs.~\cite{Beuf:2022ndu,Beuf:2021qqa,Beuf:2021srj}.

\section{Usage}
\label{sec:usage}
The most recent version of the code is available on GitHub~\cite{githubcode}, and the published version on Zenodo~\cite{zenodocode}.     Additional examples are provided in the README file in the code repository.

\subsection{Proton structure functions}

The NLO DIS cross section computation is implemented in the \code{NLODIS} class. It requires the BK-evolved dipole-target scattering amplitude that the user has to provide. This has to be provided as a class inherited from the virtual class \code{Dipole} implementing the function
\begin{cppcode}
double DipoleAmplitude(double r, double Y) const
\end{cppcode}
that computes the dipole amplitude $N_{01}$.
Here \code{r}$=|\xt_0-\xt_1|$ is the dipole size in $\mathrm{GeV}^{-1}$, and \code{Y} is the evolution rapidity. 
As an example implementation, the code package provides a class \code{BKDipole} that is based on the code published as a part of the  NLO fit~\cite{Casuga:2025etc}. In particular, it directly supports datafiles available in Ref.~\cite{samples_zenodo}.

Because the total DIS cross section is only weakly sensitive to the target geometry, the code assumes that the impact parameter dependence can be factorized out and the dipole amplitude only depends on the size of the dipole: $N_{01} = N(|\xt_0-\xt_1|)$. This effectively replaces the transverse integral over the impact parameter $\bt$ by a constant factor:
\begin{equation}
\label{eq:bint_sigma02}
\int \dd[2]{\bt} \to \frac{\sigma_0}{2},
\end{equation}
where $\sigma_0/2$ is the proton transverse area determined together with the dipole amplitude when fitting the HERA data~\cite{Casuga:2025etc}. When setting up the NLO DIS calculation, the user has to provide a numerical value for this constant. 
The factorized $\bt$ dependence is not accurate for nuclear targets, where the spatial structure of the nucleus plays a significant role.  A practical approach for computing nuclear DIS structure functions within this framework is described in Sec.~\ref{sec:nuclei}.

The program is written in C++ and can be compiled using CMake. As a dependency, the GNU Scientific Library (GSL)~\cite{Galassi:2002gnu} has to be available in the system. Multidimensional numerical integrals are performed using the Vegas algorithm implemented in the Cuba library~\cite{Hahn:2004fe} which is shipped as a part of this program and automatically compiled. The code can be built as follows: 
\begin{cli}
mkdir build;
cd build;
cmake ..
make
\end{cli}
Complete API documentation is generated by running
\begin{cli}
doxygen 
\end{cli}
in the \code{docs} directory.

Unit tests can be run as
\begin{cli}
./bin/nlodis_tests
\end{cli}
in the \texttt{build} directory. The example program (see \texttt{src/main.cpp}) is run as
\begin{cli}
./bin/nlodis
\end{cli}

In addition to the dipole-target amplitude, the user can also control various other options stored in the \code{NLODIS::config} struct.
The following example code illustrates the most important parameters, see full API documentation of the NLODIS class for details:
\begin{cppcode}

#include "dipole/bkdipole/bkdipole.hpp"
#include "nlodis.hpp"
#include "qcd.hpp"
#include <iostream>

int main()
{
  NLODIS dis;
  // Read in dipole amplitude
  // This datafile can be downloaded from 
  // https://doi.org/10.5281/zenodo.15552940.
 dis.SetDipole(std::make_unique<BKDipole>("nlobkdatafiles/zenodo.15552940/balsd/bk_map.dat"));
 // Perform NLO calculation
  dis.SetOrder(Order::NLO);

 // Set other parameters according to the applied fit, see arXiv:2506.00487 table 1

  // Running coupling scale
  dis.SetRunningCouplingC2(1.74); 
  // The distance scale is set by the smallest dipole size
  dis.SetRunningCouplingScheme(RunningCouplingScheme::SMALLEST);
  // Smooth IR behavior
  dis.SetRunningCouplingIRScheme(RunningCouplingIRScheme::SMOOTH);
  
  // Charm quark mass
  dis.SetQuarkMass(Quark::Type::C, 1.24);
  // Light quark mass is exactly 0 in arXiv:2506.00487
  // Use a finite value here for numerical stability
  dis.SetQuarkMass(Quark::Type::LIGHT, 0.005);
  
  // Proton transverse area
  dis.SetProtonTransverseArea(9.08, Unit::MB);
  
  // Print parameters to stdout
  dis.PrintConfiguration(); 

  // Compute F2
  double Q2=8.5;
  double xbj=1e-3;
  double F2 = dis.F2(Q2,xbj);

  std::cout << "F_2(Q^2=" << Q2 << " GeV^2, xbj=" << xbj << ") = " << F2 << std::endl;

  return 0;
}
\end{cppcode}
This minimal working example is available in \code{src/simple_example.cpp} and is also compiled automatically.
In this particular case, the output would be
\begin{cli}
=== NLODIS Configuration Summary ===
Order: NLO
Subtraction Scheme: Unsubtracted (UNSUB)
Nc Scheme: Finite Nc
Running Coupling Scale: Smallest dipole
Heavy quark x scheme: MassIndependentX, x = x_Bj = Q^2/(Q^2+W^2)
IR Freezing Scheme: Smooth
Maximum dipole size (maxr): 30 GeV^-1
Running coupling C^2 factor: 1.74
Non-perturbative scale (Q0^2): 1 GeV^2
Proton transverse area: 23.3192 GeV^-2 = 9.08067 mb
Quark flavors and masses:
  light, m=0.005 GeV (e=0.816497)
  c, m=1.24 GeV (e=0.666667)
Active flavors for running coupling: determined from quark list
Dipole: Data read from file nlobkdatafiles/zenodo.15552940/balsd/bk_map.dat, minr: 1e-06 GeV^(-1), maxr: 27.5255 GeV^(-1), rpoints: 200 initial rapidity 0, maximum rapidity 15.44 Q_{s,0}^2(initial rapidity) = 0.223376 GeV^2, Q_{s,0}^2(Y=ln 1/0.01) = 0.195612 GeV^2 [ N(r^2=2/Q_s^2) = 0.393469]
Cuba integration method: vegas, maxeval 2000000, relaccuracy 0.001 cores not set (using default)
===================================

F_2(Q^2=8.5 GeV^2, xbj=0.001) = 0.875895
\end{cli}
Using the default Monte Carlo integration settings, the runtime for this example program is approximately 2 minutes on a single CPU core of a typical laptop (tested on a 2022 MacBook Pro).

Here, the parameter $C^2$ (set by \code{NLODIS::SetRunningCouplingC2}) controls the running coupling scale in the coordinate space: the coupling is evaluated at the scale $4C^2/r^2$.  By default all dimensionful variables in the code are in powers of GeV (e.g. $[\text{distance}]=\mathrm{GeV}^{-1}$). When setting the proton transverse area, the user can specify that the value is in millibarns by using the optional second  argument. Note that if one uses the NLO dipoles fitted in Ref.~\cite{samples_zenodo}, one has to  manually set the parameters to be the same as in the fit, the datafiles only provide the dipole amplitude.

The structure functions $F_2$, $F_L$ and $F_T$ can be calculated as follows
\begin{cppcode}
double Q2 =  10.0; // In GeV^2
double xbj = 0.001; // Bjorken-x
cout << "F_2 = " << dis.F2(Q2, xbj) << ", F_T = " << dis.FT(Q2, xbj) << ", F_L = " << dis.FL(Q2, xbj) << endl;
\end{cppcode}
Note that internally $F_2$ is calculated by summing  $F_T$ and $F_L$, and in performance-critical applications where both $F_2$ and $F_L$ are needed (e.g. when computing the reduced cross section), one should  compute $F_T$ and $F_L$ separately, and then evaluate $F_2=F_L+F_T$.

Multidimensional integrations are performed using Cuba. Options controlling the applied integration algorithm, the maximum number of Monte Carlo integration points and the accuracy goal and stored in the struct \code{NLODIS::cuba_config} and can be controlled by the methods 
\begin{cppcode}
void NLODIS::SetMCIntegrationPoints(const int points)
void NLODIS::SetMCIntegrationMethod(const std::string& method)
\end{cppcode}
The number of Monte Carlo integration points required for precise results depends on kinematics, quark masses and other parameters. As such, the user should always check that the obtained results are independent of the number of Monte Carlo integration points used.

\subsection{Nuclear structure functions}
\label{sec:nuclei}

The current implementation does not support impact parameter $\bt$ dependent dipole amplitudes. Results for the nuclear structure functions can however be computed if one assumes that the BK evolution of the dipole-nucleus amplitude can be calculated independently for each $\bt$. This approximation has the advantage that one does not need to specify a model to regulate confinement scale effects (see e.g.~\cite{Berger:2010sh,Mantysaari:2024zxq}). 

In the optical Glauber model based approach suggested in Ref.~\cite{Lappi:2013zma}, which employs the $\bt$-independent BK evolution approximation, one replaces the parameter $(Q_{s,0}^2)^\gamma$ controlling the proton saturation scale at the initial condition of the BK evolution by a $\bt$-dependent nuclear saturation scale:
\begin{equation}
    (Q_{s,A}^2)^\gamma = A T_A(\bt) \frac{\sigma_0}{2} (Q_{s,p}^2)^\gamma.
\end{equation}
Here $\gamma$ is the anomalous dimension at the initial condition which is determined, together with the proton transverse area and other parameters describing the initial dipole-proton scattering amplitude, by fitting to the HERA data. 

The nuclear structure functions can then be calculated by first computing the virtual photon-nucleus cross section separately at each impact parameter, $\dd \sigma^{\gamma^* A}/\dd[2]\bt$ using this program, and then separately performing the integral over the impact parameter. For example in Ref.~\cite{Casuga:2023dcf}  this approach was used to compute predictions for the nuclear modification factor for the DIS structure functions in EIC kinematics. For a more detailed discussion about this procedure, the reader is referred to Ref.~\cite{Lappi:2013zma}.

\section{Cross section in a form suitable for numerical evaluation}
\label{sec:explicit_xs}

The code calculates the three different contributions shown in Eq.~\eqref{eq:nloxs}, $\sigmalt{\textrm{IC}}$,  $\sigmaltdip$ and  $\sigmaltqgu$, separately for transverse and longitudinal virtual photon.

\subsection{Lowest order contribution}

The lowest order contribution  $\sigmalt{\textrm{IC}}$ corresponds to the case where a quark-antiquark system interacts with the shockwave, and the dipole-target scattering amplitude is evaluated at the initial condition. Furthermore, no loop corrections to the virtual photon wave function are included. Note that the cross section at leading logarithmic accuracy (corresponding to leading order DIS fits performed e.g. in Refs.~\cite{Albacete:2010sy,Lappi:2013zma,Casuga:2023dcf}) can be obtained by evaluating the Wilson line correlator at $Y=\ln(1/\xbj)$, instead of setting $Y=Y_0$ which is the case when evaluating $\sigma^\mathrm{IC}$.

The cross section for a longitudinally polarized virtual photon can be written as
\begin{equation}
    \sigma_{L}^\mathrm{IC} = 4 \nc \aem 4Q^2 \sum e_f^2 \int_0^1 \dd z \int \frac{\dd[2]\bt}{2\pi} \frac{\dd[2]\xt_{01}}{2\pi} Q^2[z(1-z)]^2 \left[ K_0\left(\kappa_z |\xt_{01}|\right)\right]^2 N_{01}.
\end{equation}
Here  
\begin{equation}
    \kappa_z^2=Q^2z(1-z)+m^2.
\end{equation}
We use the notation $\xt_{ij}=\xt_i-\xt_j$, and $f$ is the quark flavor.
Note that in principle the cross section is a function of the kinematical variables $\xbj$ and $Q^2$, but as the BK evolution is formally higher order in $\as$, the lowest-order contribution is $\xbj$-independent.
In this work the quark mass is denoted by $m$ and it implicitly depends on the quark flavor $f$. Furthermore, $e_f$ is the fractional charge of the quark.

Similarly the lowest order transverse cross section reads
\begin{equation}
    \sigma_{T}^\mathrm{IC} = 4 \nc \aem \sum e_f^2 \int_0^1 \dd z \int \frac{\dd[2]\bt}{2\pi} \frac{\dd[2]\xt_{01}}{2\pi}  \Big\{ [z^2+(1-z)^2] \kappa_z^2 K_1(\kappa_z \xt_{01}) + m^2 K_0(\kappa_z |\xt_{01}|)\Big\} N_{01}
\end{equation}

In the numerical implementation, the lowest order contribution is calculated in 
\begin{cppcode}
double NLODIS::Photon_proton_cross_section_LO_d2b(double Q2, double xbj, Polarization pol)
\end{cppcode}
It calculates $\dd{\sigma^\mathrm{LO}_{L,T}}/\dd[2]{\bt}$.
Here \code{Polarization} is an \code{enum class} with possible values \code{T} and \code{L} referring to transverse and longitudinal photon polarization states, respectively.
 As we assume an impact-parameter independent dipole, there is no dependence on the dipole orientation and one overall angular integral (orientation of $\xt_{01}$) is performed analytically. When calculating the structure functions in 
\begin{cppcode}
double NLODIS::F2(double Q2, double xbj)
double NLODIS::FL(double Q2, double xbj)
\end{cppcode}
the result is multiplied by the proton transverse area, see Eq.~\eqref{eq:bint_sigma02} and the  discussion in Sec.~\ref{sec:nuclei}.

\subsection{Longitudinal polarization, dipole contribution}
\label{sec:dip_L}
For a longitudinally polarized virtual photon, the next-to-leading order dipole contribution is~\cite{Beuf:2021qqa}
\begin{multline}
\label{eq:NLO_dip_L}
    \sigma_L^\mathrm{dip} =     4\nc\aem4Q^2\sum_{f}e_f^2\int_{0}^{1} \dd z  \int \frac{\dd[2]\bt}{2\pi} \int \frac{\dd[2]\xt_{01}}{2\pi}  [z(1-z)]^2 \\
\times  \Biggl \{ \left (\frac{\alpha_s\cf}{\pi}\right ) \biggl[ \frac{5}{2} - \frac{\pi^2}{3} + \log^2\left (\frac{z}{1-z}\right )   + \Omega_{\mathcal{V}}^L(\gamma;z) + L(\gamma;z) \biggr]  [K_0\left (\vert \kappa_z \xt_{01}\vert \right)
 ]^2\\
 + \left (\frac{\alpha_s\cf}{\pi}\right )K_0\left (\kappa_z \vert \xt_{01}\vert
 \right )\widetilde{\mathcal{I}}_{\mathcal{V}}(z,\xt_{01})  \Biggr \}N_{01}.
\end{multline}
The strong coupling is evaluated at the only available transverse length scale $\as=\as(|\xt_{01}|)$; see Sec.~\ref{sec:coupling} for a discussion of the possible strong coupling choices.
In the code, the contribution~\eqref{eq:NLO_dip_L} is calculated in 
\begin{cppcode}
double NLODIS::Sigma_dip_d2b(double Q2, double xbj, Polarization pol))
\end{cppcode}
The special functions that appear in Eq.~\eqref{eq:NLO_dip_L} are defined in Appendix~\ref{appendix:special_fun_L_dip}.

The function $\widetilde{\mathcal{I}}_{\mathcal{V}}$ contains two terms with one and two extra numerical integrals, see Eqs.~\eqref{eq:IVsumFT}, \eqref{eq:J1int} and \eqref{eq:J2int}. In the numerical evaluation the different contributions are always grouped based on the dimension of the numerical integration, and the  integration variables from these special functions are included in the Monte Carlo integral that also integrates over $z$ and $|\xt_{01}|$ (and other transverse coordinates and momentum fractions when evaluating the $\sigmaltqgu$ contribution). Explicitly, we rewrite Eq.~\eqref{eq:NLO_dip_L} as
\begin{equation}
\label{eq:dip_L_decomposition}
    \sigma_{L}^\mathrm{dip} = \sigma_{L0}^\mathrm{dip} + \sigma_{L\ (a)+(b)}^\mathrm{dip} + \sigma_{L\ (c)+(d)}^\mathrm{dip}.
\end{equation}
Here the subscripts $(a)+(b)$ and $(c)+(d)$ refer to the notation used in \cite{Beuf:2021qqa}. These three terms have a different number of numerical integrations. Explicit expressions are given below. First, the term where the numerical integration is 2 dimensional (as we replace $\int \dd[2]{\bt}$ by a constant) reads
\begin{multline}
\label{eq:NLO_dip_L_1}
    \sigma_{L0}^\mathrm{dip} = 
    4\nc\aem4Q^2\sum_{f}e_f^2\int_{0}^{1} \dd z  \int \frac{\dd[2]\bt}{2\pi} \int \frac{\dd[2]\xt_{01}}{2\pi}  [z(1-z)]^2 \\
\times   \left (\frac{\alpha_s\cf}{\pi}\right ) \biggl [ \frac{5}{2} - \frac{\pi^2}{3} + \log^2\left (\frac{z}{1-z}\right )   + \Omega_{\mathcal{V}}^L(\gamma;z) + L(\gamma;z) \biggr ]\biggl [K_0\left (\kappa_z\vert \xt_{01}\vert
\right )\biggr ]^2 N_{01}.
\end{multline}
This integrand is implemented in 
\begin{cppcode}
double ILdip_massive_Omega_L_Const(double Q2, double z, double r, double mf)
\end{cppcode}
For all three terms discussed here, the overall prefactors and the evaluation of the dipole amplitude $N_{01}$ are included in the Cuba integration wrapper
\begin{cppcode}
int integrand_dip_massive(const int *ndim, const double x[], const int *ncomp, double *f, void *userdata)
\end{cppcode}

The second contribution $\sigma_{L\ (a)+(b)}^\mathrm{dip}$ has a three dimensional numerical integration:
\begin{multline}
    \sigma_{L\ (a)+(b)}^\mathrm{dip} = 
     4\nc\aem4Q^2\sum_{f}e_f^2\int_{0}^{1} \dd z  \int \frac{\dd[2]\bt}{2\pi} \int \frac{\dd[2]\xt_{01}}{2\pi}  [z(1-z)]^2 \\
     \times \left (\frac{\alpha_s\cf}{\pi}\right )K_0\left (\kappa_z \vert \xt_{01}\vert
     \right ) \widetilde{\mathcal{I}}_{\mathcal{V}_{(a)+(b)}}(z,\xt_{01})  N_{01}. 
    \end{multline}
Here $\widetilde{\mathcal{I}}_{\mathcal{V}_{(a)+(b)}}(z,\xt_{01})$, given in Eq.~\eqref{eq:J1int}, includes one additional numerical integration over $\xi$, which is included in the Monte Carlo integration. The integrand is implemented in
\begin{cppcode}
double ILdip_massive_Iab(double Q2, double z1, double r, double mf, double xi)
\end{cppcode}

Similarly the last term in Eq.~\eqref{eq:dip_L_decomposition} reads
\begin{multline}
\label{eq:NLO_dip_L_2}
   \sigma_{L\ (c)+(d)}^\mathrm{dip} = 
    4\nc\aem4Q^2\sum_{f}e_f^2\int_{0}^{1} \dd z  \int \frac{\dd[2]\bt}{2\pi} \int \frac{\dd[2]\xt_{01}}{2\pi}  [z(1-z)]^2 \\
    \times \left (\frac{\alpha_s\cf}{\pi}\right )K_0\left (\kappa_z\vert \xt_{01}\vert
    \right )\widetilde{\mathcal{I}}_{\mathcal{V}_{(c)+(d)}}(z,\xt_{01})  N_{01}. 
    \end{multline}
Here $\widetilde{\mathcal{I}}_{\mathcal{V}_{(c)+(d)}}(z,\xt_{01})$, given in Eq.~\eqref{eq:J2int}, includes two additional numerical integrations over $\xi$ and $x$ (not to be confused with Bjorken-$\xbj$), which are included in the Monte Carlo integration that becomes four dimensional. The integrand is implemented in
\begin{cppcode}
double ILdip_massive_Icd(double Q2, double z1, double r, double mf, double xi, double x) 
    \end{cppcode}

\subsection{Transverse photon, dipole contribution}
\label{sec:dip_T}
For a transversally polarized virtual photon, the dipole term can be written as~\cite{Beuf:2022ndu}
\begin{equation}
\label{eq:sigma_dip_T}
\begin{split}
\sigma^\mathrm{dip}_{T}  & = 4\nc\aem\sum_{f}e_f^2 \int_{0}^{1} \dd z \int \frac{\dd[2]{\xt_{01}}}{2\pi}\frac{\dd[2]{\bt}}{2\pi}  \Biggl \{ \biggl [z^2 + (1-z)^2 \biggr ]
\biggl [\kaz K_{1}\left (\vert \xt_{01}\vert \kaz \right )\biggr ]^2 +  m^2\biggl [K_{0}\left (\vert \xt_{01}\vert \kaz\right )\biggr ]^2 \\
& + \left (\frac{\alpha_s\cf}{\pi}\right )\biggl \{\biggl [z^2 + (1-z)^2 \biggr ]\kaz K_1(\vert \xt_{01}\vert \kaz)f_{\mathcal{V}} + \frac{(2z-1)}{2} \kaz K_1(\vert \xt_{01}\vert \kaz) f_{\mathcal{N}}\\
& + m^2 K_0(\vert \xt_{01}\vert \kaz) f_{VMS} \biggr \}\Biggr \}N_{01} .
\end{split}
\end{equation}
Here again $\kappa_z^2 = z(1-z)Q^2+m^2$, and the strong coupling is  evaluated at $\as = \as(|\xt_{01}|)$.

In the code, this contribution is calculated by the same function 
\begin{cppcode}
double NLODIS::Sigma_dip_d2b(double Q2, double xbj, Polarization pol))
\end{cppcode}
as in the case of longitudinally polarized photon, see discussion  in Sec.~\ref{sec:dip_L}. The special functions that appear in Eq.~\eqref{eq:sigma_dip_T} are defined in Appendix~\ref{appendix:special_fun_T_dip}. 

As the special functions have a different number of numerical integrations, we rewrite Eq.~\eqref{eq:sigma_dip_T} by dividing it into different parts according to the dimension of the numerical integral 
\begin{align}
\sigma_T^\mathrm{dip}
=
\sigma_{T0}^\mathrm{dip}
+\sigma_{T1}^\mathrm{dip}
+\sigma_{T2}^\mathrm{dip}.
\end{align}
 In the first term the numerical integral is two dimensional (integration variables $|\xt_{01}|$ and $z$) and it reads
\begin{multline}
\sigma_{T0}^\mathrm{dip}
=
4 N_c \alpha_{\mathrm{em}}
\sum e_f^2
\int_0^1 \dd z
\int \frac{\dd[2]{\xt_{01}}}{2\pi}
\int \frac{\dd[2] \bt}{2\pi}\,
\left(\frac{\as \cf}{\pi}\right)
\\
 \times
\Bigg[
\left(\kappa_z K_1(|\mathbf{x}_{01}|\kappa_z)\right)^2
\Bigg\{
[z^2+(1-z)^2]
\left(
\frac{5}{2}
-\frac{\pi^2}{3}
+\ln^2\left(\frac{z}{1-z}\right)
+\Omega_\mathcal{V}^T(\gamma;z)
+L(\gamma;z)
\right)
+\frac{2z-1}{2}\,\Omega_{\mathcal{N}}^T
\Bigg\}
\\
+m^2
\left(K_0(\kappa_z|\mathbf{x}_{01}|)\right)^2
\left(
3
-\frac{\pi^2}{3}
+\ln^2\left(\frac{z}{1-z}\right)
+\Omega_\mathcal{V}^T(\gamma;z)
+L(\gamma;z)
\right)
\Bigg]
\,N_{01}.
\end{multline}
In the code this integrand is implemented in
\begin{cppcode}
double ITdip_massive_0(double Q2, double z1, double x01sq, double mf)
\end{cppcode}
The special functions $\Omega_\mathcal{V}^T(\gamma;z)$ and $\Omega_\mathcal{N}^T$ that appear above are given in Appendix~\ref{appendix:special_fun_T_dip}, Eqs.~\eqref{eq:SigmaOmega} and~\eqref{eq:degOmegaN}. The function $L(\gamma;z)$ is given in Eq.~\eqref{eq:Lfunction}.

The second term  has an integration dimension 3: in addition to $|\xt_{01}|$ and $z$, we integrate over $\xi$ in the special functions  $\widetilde I_{V,1}^T$, Eq.~\eqref{eq:IVT1}, and $\widetilde I_{VMS,1}^T$, Eq.~\eqref{eq:IVMST1}
\begin{align}
\sigma_{T1}^\mathrm{dip}
&=
4 N_c \alpha_{\mathrm{em}}
\sum e_f^2
\int_0^1 \dd z
\int  \frac{\dd[2] \mathbf{x}_{01}}{2\pi}
\int  \frac{\dd[2] \bt}{2\pi}
\left(\frac{\as \cf}{\pi}\right)
\nonumber\\
&\quad \times
\Bigg[
K_1(|\mathbf{x}_{01}|\kappa_z)\,\kappa_z
\Big\{
[z^2+(1-z)^2]\,
\widetilde I_{V,1}^T
\Big\}
+
m^2 K_0(|\mathbf{x}_{01}|\kappa_z)\,
\widetilde I_{VMS,1}^T
\Bigg]
\,N_{01}.
\end{align}
 
In the third term the integration dimension is 4: in addition to $|\xt_{01}|$ and $z$, there are two numerical integrations in the special functions $\widetilde I_N^T$, Eq.~\eqref{eq:INT}, $\widetilde I_{V,2}^T$, Eq.~\eqref{eq:IVT2}, and $\tilde I_{VMS2}^T$, Eq.~\eqref{eq:IVMST2}. This contribution reads
\begin{multline}
    \sigma^\mathrm{dip}_{T2}
= 4 N_c \alpha_{\mathrm{em}}
\sum e_f^{\,2}
\int_0^1 \dd{z} 
\int \frac{\dd[2]{\xt_{01}}}{2\pi}
\int \frac{\dd[2]{\bt}}{2\pi}
\left(\frac{\as \cf}{\pi}\right)
\Bigg[
\kappa_zK_1(|\mathbf{x}_{01}| \kappa_z)
\left\{
\left[z^2 + (1-z)^2\right] \widetilde{I}^{T}_{V,2} \right. \\
\left.
+ \frac{2z - 1}{2}\,\widetilde{I}^{T}_{{N}}
\right\} 
+ m^{\,2} K_0(|\mathbf{x}_{01}|\kappa_z)\,\widetilde{I}^T_{VMS,2}
\Bigg]\,
N_{01}.
\end{multline}

Similarly as in the case of longitudinal polarization, these integrands are called from the function \code{integrand_dip_massive} that provides them in a form suitable for the Cuba integration routines and evaluates the dipole-target scattering amplitude $N_{01}$. In the code, the integrals are implemented in
\begin{cppcode}
double ITdip_massive_1(double Q2, double z1, double x01sq, double mf, double xi)
double ITdip_massive_2(double Q2, double z1, double x01sq, double mf, double y_chi, double y_u) 
\end{cppcode}

\subsection{Longitudinal photon, \texorpdfstring{$q \bar q g$}{qqg} contribution}
\label{sec:L_qqg}
In the exponential ultraviolet subtraction scheme used in this work, the contribution from diagrams where the $|q\bar qg\rangle$ state interacts with the shockwave reads in the longitudinal photon case as~\cite{Beuf:2021qqa}
\begin{multline}
\label{eq:sigma_qg_L}
\sigma_L^{q\bar q g}    =  4\nc\aem 4Q^2\sum_{f}e_f^2
\int \frac{\dd[2]{\bt}}{2\pi} \int \frac{\dd[2]\xt_{01}}{2\pi} \frac{\dd[2] \xt_{02}}{2\pi}
\int_0^{1-\zmin} \dd {z_1} \int_{\zmin}^{1-z_1} \frac{\dd{z_2}}{z_2} 
\left (\frac{\alpha_s\cf}{\pi}\right )\\
\times \Biggl \{z_1^2\biggl [2z_0(z_0 + z_2) + (z_2)^2 \biggr ]   \biggl \{   \frac{\vert\xt_{20}\vert^2}{64} \left[\mathcal{G}_{(k)}^{(1;2}\right] ^2 N_{012}  - \frac{e^{-\vert\xt_{20}\vert^2/(\vert\xt_{01}\vert^2 e^{\gamma_E})}}{\vert\xt_{20}\vert^2} \biggl [K_0 \left (\vert \xt_{01}\vert \sqrt{\overline{Q}^2_{(k)} + m^2} \right )\biggr ]^2   N_{01}  \biggr \} \\
 + (z_0)^2\biggl [2z_1(z_1 + z_2) + (z_2)^2 \biggr ] 
   \biggl \{   \frac{\vert\xt_{21}\vert^2}{64}\left[\mathcal{G}_{(l)}^{(1;2)}\right]^2  N_{012}  - \frac{e^{-\vert\xt_{21}\vert^2/(\vert\xt_{01}\vert^2 e^{\gamma_E})}}{\vert\xt_{21}\vert^2} \biggl [K_0 \left (\vert \xt_{01}\vert \sqrt{\overline{Q}^2_{(l)} + m^2} \right )\biggr ]^2   N_{01}  \biggr \} \\
 - \frac{z_0z_1}{32} \biggl [z_1(z_0 + z_2) + z_0(z_1 + z_2) \biggr ] (\xt_{20}\cdot \xt_{21}) [\mathcal{G}_{(k)}^{(1;2)}][\mathcal{G}_{(l)}^{(1;2)}]N_{012}\\
  + \frac{m^2}{16}(z_2)^4 \Biggl [\frac{z_1}{(z_0 + z_2)} [\mathcal{G}_{(k)}^{(1;1)}]  - \frac{z_0}{(z_1 + z_2)}[\mathcal{G}_{(l)}^{(1;1)}] \Biggr ]^2 
N_{012}\Biggr \}.
\end{multline}
Here $z_0 = 1-z_1-z_2$, $\overline{Q}^2_{(k)}=z_1(1-z_1)Q^2$ and $\overline{Q}^2_{(l)}=z_0(1-z_0)Q^2$.
Following Ref.~\cite{Beuf:2020dxl}, the lower limit of the $z_2$ integral is
\begin{equation}
    \zmin = \frac{Q_0^2}{W^2}.
\end{equation}
Here $Q_0^2$ is a non-perturbative target energy scale.

Unlike in the dipole contributions, there are now three different transverse length scales available in the $q\bar q g$ contributions when evaluating the strong coupling: $|\xt_{01}|,|\xt_{02}|$ and $|\xt_{12}|$. If the running coupling scheme is set as
\begin{cppcode}
NLODIS::SetRunningCouplingScheme(RunningCouplingScheme::PARENT)
\end{cppcode}
then the coupling is evaluated at $\as=\as(|\xt_{01}|)$. The other option is to select
\begin{cppcode}
NLODIS::SetRunningCouplingScheme(RunningCouplingScheme::SMALLEST)
\end{cppcode}
in which case the smallest length scale controls the scale of the coupling
\begin{equation}
\as=\as\left(\min\{|\xt_{01}|,|\xt_{02}|,|\xt_{12}|\}\right).
\end{equation}

In our numerical implementation we manipulate the generalized Bessel function defined in \cite{Beuf:2021qqa} as 
\begin{equation}
\label{eq:gen_bessel}
    \mathcal{G}_{(x)}^{(a;b)} = \int_{0}^{\infty} \dd u \int_{0}^{u/\omega_{(x)}} \dd t \frac{1}{u^a t^b} g_{(x)}(t,u)
 \end{equation}
 where $(x)=(l)$ or $(k)$,  and
 \begin{equation}
     g_{(x)}(t,u) = e^{-u[\overline{Q}^2_{(x)} + m^2]} e^{-\frac{\vert\xt_{3;(x)}\vert^2}{4u}}  e^{-t\omega_{(x)}\lambda_{(x)}m^2}e^{-\frac{\vert\xt_{2;(x)}\vert^2}{4t}}.
 \end{equation}
 Here $\lambda_{(k)}=z_1z_2/z_0$, $\lambda_{(l)}=z_0z_2/z_1$, $\omega_{(k)}=z_0z_2/[z_1(z_0+z_2)^2]$ and $\omega_{(l)}=z_1z_2/[z_0(z_1+z_2)^2]$. The transverse coordinates are
 \begin{align}
    \xt_{2;(k)}&=\xt_{20}, \\
    \xt_{2;(l)}&=\xt_{21}, \\
     \xt_{3;(j)}&=\left(\frac{z_0 \xt_0 + z_2 \xt_2}{z_0+z_2}\right) - \xt_1,\\
     \xt_{3;(k)}&=\xt_0 - \left(\frac{z_1 \xt_1 + z_2 \xt_2}{z_0+z_2}\right),
 \end{align}
 which gives
 \begin{align}
\left| \mathbf{x}_{3;(k)} \right|^2 
&= \frac{z_0^2}{(z_0 + z_2)^2} \mathbf{x}_{20}^2
   + \mathbf{x}_{21}^2
   - 2 \frac{z_0}{z_0 + z_2} \, \mathbf{x}_{20} \cdot \mathbf{x}_{21} \\
\left| \mathbf{x}_{3;(l)} \right|^2 
&= \mathbf{x}_{20}^2
   + \frac{z_1^2}{(z_1 + z_2)^2} \mathbf{x}_{21}^2
   - 2 \frac{z_1}{z_1 + z_2} \, \mathbf{x}_{20} \cdot \mathbf{x}_{21}.
\end{align}

We first perform a change of variables $t \to y = t \omega_{(x)}/u$, after which the $u$-integral can be done analytically. This gives
\begin{multline}
    \label{eq:gen_bessel_integrated}
\mathcal{G}^{(a;b)}_{(x)}
=
\int_{0}^{1} \frac{\dd y}{y^{\frac{1}{2}(2-a-b)}}
\, 2^{a+b-1} \, \omega_{(x)}^{\, b-1}
\left(
\frac{ \lambda_{(x)} m^{2} + \overline{Q}^{\,2}_{(x)} + m^{2} }
     { y \left| \mathbf{x}_{3;(x)} \right|^{2}
       + \omega_{(x)} \left| \mathbf{x}_{2;(x)} \right|^{2} }
\right)^{\frac{1}{2}(a+b-2)} \\
\times
K_{a+b-2}\left(
\sqrt{
\frac{1}{y}
\left( \lambda_{(x)} m^{2} + \overline{Q}^{\,2}_{(x)} + m^{2} \right)
\left(
y \left| \mathbf{x}_{3;(x)} \right|^{2}
+ \omega_{(x)} \left| \mathbf{x}_{2;(x)} \right|^{2}
\right)
}
\right)
=\int_0^1 \dd{y} g^{(a;b)}_{(x)}(y),
\end{multline}
where
\begin{multline}
    g^{(a;b)}_{(x)}(y) = \frac{1}{y^{\frac{1}{2}(2-a+b)}} 2^{a+b-1} \omega_{(\textrm{x})}^{b-1} \left( \frac{y |\xt_{3;(\textrm{x})}|^2 +\omega_{(\textrm{x})}|\xt_{2;(\textrm{x})}|^2 }{y  \lambda_{(\textrm{x})} m^2 +  \overline Q^2_{(\textrm{x})}+m^2} \right)^{\frac{1}{2}(2-a-b)}\\
\times     K_{a+b-2} \left(\sqrt{\frac{1}{y} \left( y \lambda_{(\textrm{x})} m^2 +  \overline Q^2_{(\textrm{x})}+m^2\right) \left(y |\xt_{3;(\textrm{x})}|^2 +\omega_{(\textrm{x})}|\xt_{2;(\textrm{x})}|^2  \right)} \right).
\end{multline}
Furthermore in the limit $\lambda_{(x)}\to 0$, this integral can be evaluated analytically when $b=2$. The result is
\begin{align}
\mathcal{G}^{(a;2)}_{(x)}\left(\lambda_{(x)} \to 0\right)
&=
\frac{2^{2+a}}{\left| \mathbf{x}_{2;(x)} \right|^{2}}
\left(
\frac{\overline{Q}^{2}_{(x)} + m^{2}}
     {\left| \mathbf{x}_{3;(x)} \right|^{2}
      + \omega_{(x)} \left| \mathbf{x}_{2;(x)} \right|^{2}}
\right)^{\frac{a-1}{2}} \\
&\times K_{a-1}\left(
\sqrt{
\left(\overline{Q}^{2}_{(x)} + m^{2}\right)
\left(
\left| \mathbf{x}_{3;(x)} \right|^{2}
+ \omega_{(x)} \left| \mathbf{x}_{2;(x)} \right|^{2}
\right)
}
\right).
\end{align}
In the numerical integrations, the integration variable $y$ is included in the phase space of the Monte Carlo integration.
Consequently, individual terms in Eq.~\eqref{eq:sigma_qg_L} can be divergent as they contain terms $\xt_{2;(x)}^{-2}$. These divergences only cancel in the sum, which is numerically challenging due to the different number of integration variables in different terms. To address this numerical challenge, we again divide terms to separate blocks according to the dimension of the numerical integral, but render each term finite by performing a replacement using the identity
\begin{multline}
\label{eq:L_G_subst}
    \mathcal{G}_{(x)}^{(a;2)} =
  \mathcal{G}_{(x)}^{(a;2)}(\lambda=0) 
+
  \qty[  \mathcal{G}_{(x)}^{(a;2)} - \mathcal{G}_{(x)}^{(a;2)}(\lambda=0) ]
\\
=
   \frac{2^{2+a}}{\left| \mathbf{x}_{2;(x)} \right|^{2}}
\left(
\frac{\overline{Q}^{2}_{(x)} + m^{2}}
     {\left| \mathbf{x}_{3;(x)} \right|^{2}
      + \omega_{(x)} \left| \mathbf{x}_{2;(x)} \right|^{2}}
\right)^{\frac{a-1}{2}} K_{a-1}\left(
\sqrt{
\left(\overline{Q}^{2}_{(x)} + m^{2}\right)
\left(
\left| \mathbf{x}_{3;(x)} \right|^{2}
+ \omega_{(x)} \left| \mathbf{x}_{2;(x)} \right|^{2}
\right)
}
\right)
\\
+
\int_0^1 \dd{y}
\overline g^{(a;2)}_{(x)}(y)
\end{multline}
where
\begin{multline}  
\overline g^{(a;b)}_{(x)}(y)
=g^{(a;b)}_{(x)}(y) - g^{(a;b)}_{(x)}(y; \lambda_{(\textrm{x})}  \to 0)
=\frac{1}{y^{\frac{1}{2}(2-a+b)}} 2^{a+b-1} \omega_{(\textrm{x})}^{b-1}\\
\times \left\{
 \left( \frac{y |\xt_{3;(\textrm{x})}|^2 +\omega_{(\textrm{x})}|\xt_{2;(\textrm{x})}|^2 }{y  \lambda_{(\textrm{x})} m^2 +  \overline Q^2_{(\textrm{x})}+m^2} \right)^{-\frac{a+b-2}{2}}
K_{a+b-2} \left(\sqrt{\frac{1}{y} \left( y \lambda_{(\textrm{x})} m^2 +  \overline Q^2_{(\textrm{x})}+m^2\right) \left(y |\xt_{3;(\textrm{x})}|^2 +\omega_{(\textrm{x})}|\xt_{2;(\textrm{x})}|^2  \right)} \right)\right.
\\
 \left.-
 \left( \frac{y |\xt_{3;(\textrm{x})}|^2 +\omega_{(\textrm{x})}|\xt_{2;(\textrm{x})}|^2 }{\overline Q^2_{(\textrm{x})}+m^2} \right)^{-\frac{a+b-2}{2}}
K_{a+b-2} \left(\sqrt{\frac{1}{y} \left(   \overline Q^2_{(\textrm{x})}+m^2\right) \left(y |\xt_{3;(\textrm{x})}|^2 +\omega_{(\textrm{x})}|\xt_{2;(\textrm{x})}|^2  \right)} \right)
\right\}.
\end{multline}

In the numerical implementation, we again factor out the overall impact parameter dependence and calculate $\dd{\sigma_L^{q\bar q g}}/\dd[2]\bt$. Furthermore, there is one overall azimuthal angle that can be integrated over analytically. The remaining integration variables are $|\xt_{01}|, |\xt_{02}|, \sphericalangle(\xt_{01},\xt_{02}),z_1$ and $z_2$.
In addition to this, after the substitution~\eqref{eq:L_G_subst}, we have terms with zero, one or two integrals over $y$ from \eqref{eq:gen_bessel_integrated} in \eqref{eq:sigma_qg_L}. 
We can then write the $q\bar qg$ contribution to the DIS cross section as
\begin{align}
\label{eq:L_qqg_decomposition}
    \sigma^{q\bar q g}_L &= 4 \nc \aem 4Q^2 \sum e_f^2 \int \frac{\dd[2]{\bt}}{2\pi} \int \frac{\dd[2]\xt_{01}}{2\pi} \frac{\dd[2] \xt_{02}}{2\pi}
\int_0^{1 -\zmin}\dd {z_1} \int_{\zmin}^1 \dd{z_2} \left(\frac{\as \cf}{\pi} \right)  \\
&\times \left[  I_\mathrm{UV}^L N_{01} + I_1^L N_{012} + I_2^L N_{012} + I_3^L N_{012} \right].
\end{align}
Note that $I_2^L$ has one additional numerical integration and $I_3^L$ has two additional numerical integrations that are included in the Monte Carlo integration.

First, for the terms that do not involve any additional integrals, i.e $I_\mathrm{UV}^L$ and $I_1^L$, the integration dimension is 5. 
Explicit expressions for these terms are given in Appendix~\ref{sec:appendix_special_fun_qqg_L}, Eqs.~\eqref{eq:qqg_L_uv} and~\eqref{eq:qqg_L_I1}. These integrands are implemented in functions
\begin{cppcode}
double ILNLOqg_massive_dipole_uvsub(double Q2, double mf, double z1, double z2, double x01sq, double x02sq, double x21sq)
double ILNLOqg_massive_tripole_part_I1(double Q2, double mf, double z1, double z2, double x01sq, double x02sq, double x21sq)
\end{cppcode}

The remaining contributions $I_2^L$ and $I_3^L$ have one or two additional numerical integrations. Explicit expressions are given in Eqs.~\eqref{eq:IL2} and~\eqref{eq:IL3}, and implemented in functions
\begin{cppcode}
double ILNLOqg_massive_tripole_part_I2(double Q2, double mf, double z1, double z2, double x01sq, double x02sq, double x21sq, double y_t)
double ILNLOqg_massive_tripole_part_I3(double Q2, double mf, double z1, double z2, double x01sq, double x02sq, double x21sq, double y_t1, double y_t2) 
    \end{cppcode}
The wrapper that provides these integrands in a format suitable for Cuba integration algorithms, and evaluates the corresponding Wilson line structures, is
\begin{cppcode}
int integrand_qgunsub_massive(const int *ndim, const double x[], const int *ncomp,double *f, void *userdata)
\end{cppcode}

\subsection{Transverse photon, \texorpdfstring{$q \bar q g$}{qqg} contribution}

For brevity we do not repeat the long expressions derived in Ref.~\cite{Beuf:2022ndu} for the $q\bar qg$ contribution in the case of a transverse photon. Instead, we directly present this contribution in a form that is more suitable for numerical implementation. First, we collect the terms according to the dimension of the numerical integration and write it as
\begin{multline}
\label{eq:qqg_T_decomposition}
\sigma_T^{q\bar q g}
=
4 \nc \aem
\sum e_f^2
\int \frac{\dd[2]{\bt}}{2\pi} \frac{\dd[2]{\xt_{02}}}{2\pi}  \frac{\dd[2]{\xt_{21}}}{2\pi}
\int_0^{1-\zmin} \dd{z_1}
\int_{z_{2,\mathrm{min}}}^{1 - z_0} \frac{ \dd{z_2}}{z_2}
\left(\frac{\as \cf}{\pi}\right)\\ \times
\Bigg[ 
I_\mathrm{UV}^T N_{01}
+[I_1^T + I_{1F}^T + I_{1F_m}^T] N_{012}
+ [
I_2^T+  I_{2m}^T +I_{2F}^T + I_{2F_m}^T ]N_{012}
+ [I_3^T
+ I_{3m}^T 
  +I_{3F}^T 
 + I_{3F_m}^T
] N_{012}
\Bigg]
\end{multline}
The UV subtraction scheme-dependent term proportional to $N_{01}$, $I_\mathrm{UV}^T$, and the terms $I_1^T$, $I_{1F}^T$ and $I_{1F_m}^T$  do not involve any additional numerical integrations from special functions.
The remaining terms include one ($I_2^T, I_{2m}^T, I_{2F}^T, I_{2F_m}^T$) or two ($I_3^T, I_{3m}^T, I_{3F}^T, I_{3F_m}^T$) additional numerical integrations. Their explicit expressions are given in
Appendix~\ref{sec:appendix_special_fun_qqg_T}, Eqs.~\eqref{eq:qqg_T_uv}, \eqref{eq:qqg_T_1}, \eqref{eq:qqg_T_1F}, \eqref{eq:qqg_T_1Fm}, \eqref{eq:qqg_T_2}, \eqref{eq:qqg_T_2m}, \eqref{eq:qqg_T_2F}, \eqref{eq:qqg_T_2Fm}, \eqref{eq:qqg_T_3}, \eqref{eq:qqg_T_3m}, \eqref{eq:qqg_T_3F} and \eqref{eq:qqg_T_3Fm}.

These expressions are obtained by performing the same replacement~\eqref{eq:L_G_subst} to the generalized 
Bessel functions $\mathcal{G}_{(x)}^{(a;b)}$, that appear in the original result obtained in Ref.~\cite{Beuf:2022ndu}, as in the case of longitudinal photon, see discussion in Sec.~\ref{sec:L_qqg}.

\subsection{Running coupling}
\label{sec:coupling}
The coordinate space running coupling constant is implemented in
\begin{cppcode}
double NLODIS::Alphas(const double r) const
\end{cppcode}
The behavior of the coupling in the infrared region is controlled by\\
\code{NLODIS::config::rc_ir_scheme}.
With the \code{RunningCouplingIRScheme::FREEZE} option, maximum value allowed for the coupling is  $\alpha_{\mathrm{s,\mathrm{max}}}$. The user can set this value by calling \code{NLODIS::SetRunningCouplingMaxAlphaS(const double max_alpha_s)}. In this case the coupling reads
\begin{equation}
    \as(r) = \frac{12\pi}{(11\nc - 2\nf) \ln \frac{4C^2}{r^2 \Lambda_\mathrm{QCD}^2}}
\end{equation}
The scale of the coupling in momentum space is $4C^2/r^2$, and the user can set $C^2$ by calling \code{NLODIS::SetRunningCouplingC2(const double C2)}. The number of active quark flavors $\nf$ can be set by calling \code{NLODIS::SetActiveFlavors(const int nf)}. By default $\nf$ is computed based on the quark flavors included in the cross section calculation. Here we use $\Lambda_\mathrm{QCD}=0.241\,\mathrm{GeV}$.

The other possible option for the infrared extrapolation is \code{RunningCouplingIRScheme::SMOOTH}. In that case the coupling reads
\begin{equation}
    \as(r) = \frac{12\pi}{(11\nc - 2\nf) \ln \left[ 
    \left(\frac{\mu_0^2}{\Lambda_\mathrm{QCD}^2}\right)^{1/\zeta} + \left(\frac{4C^2}{r^2\Lambda_\mathrm{QCD}^2}\right)^{1/\zeta}
    \right]^\zeta}
\end{equation}
The parameter values $\mu_0/\Lambda_\mathrm{QCD}=2.5$ and $\zeta=0.2$ are chosen following Refs.~\cite{Beuf:2020dxl}.

\section{Conclusion}
The numerical program reported in this work enables one to calculate DIS structure functions in dipole picture at next-to-leading order accuracy. 
Depending on the required numerical accuracy and kinematical point (more Monte Carlo points are needed at higher $Q^2$ especially with small quark masses), the code typically takes $\mathcal{O}(1)$ minute on a basic laptop to compute the $F_2$ structure function at the given kinematical point using safe Monte Carlo settings. The most time consuming part is the computation of the numerical integrals using Monte Carlo routines from Vegas. In practical applications one typically has to evaluate the structure function in many different kinematical points, but as the initialization phase is fast and the memory requirements are modest, many different processes can be run simultaneously. As such, the setup can be parallelized trivially.
Consequently, the developed program is also suitable for global analyses where a large number of cross section evaluations at different kinematical points with different initial conditions for the Balitsky-Kovchegov equation are required.

\section*{Acknowledgements}
We thank C. Casuga for helpful discussions.

\paragraph{Funding information}
H.H. is supported by the Research Council of Finland (Flagship of Advanced Mathematics for Sensing Imaging and Modelling grant 359208), and the Vilho, Yrjö and Kalle Väisälä Foundation.
H.M is supported by the Research Council of Finland, the Centre of Excellence in Quark Matter, and projects 338263 and 359902, and by the European Research Council (ERC, grant agreements  No. ERC-2023-101123801 GlueSatLight and ERC-2018-ADG-835105 YoctoLHC).
J.P. is supported by the National Science Foundation under grant No.~PHY-2515057 and by the U.S. Department of Energy, Office of Science, Office of Nuclear Physics, within the framework of the Saturated Glue (SURGE) Topical Theory Collaboration.
Computing resources from CSC – IT Center for Science in Finland and the Finnish Grid and Cloud Infrastructure (persistent identifier \texttt{urn:nbn:fi: research-infras-2016072533}) were used in this work.
The content of this article does not reflect the official opinion of the European Union and responsibility for the information and views expressed therein lies entirely with the authors.

\begin{appendix}
\numberwithin{equation}{section}

\section{Explicit expressions in numerical implementation}

\subsection{Longitudinal photon, dipole contribution}
\label{appendix:special_fun_L_dip}

The special functions on the second line of Eq.~\eqref{eq:NLO_dip_L} read
\begin{multline}
\label{eq:GammaLm}
\Omega_{\mathcal{V}}^L(\gamma;z)  =   \frac{1}{2z}\biggl [\log(1-z) + \gamma \log\left (\frac{1+\gamma}{1+\gamma - 2z}\right ) \biggr ]  + \frac{1}{2(1-z)}\biggl [\log(z) + \gamma \log\left (\frac{1+\gamma}{1+\gamma - 2(1-z)}\right ) \biggr ] \\
 + \frac{1}{4z(1-z)}\left ( \gamma - 1\right )\log\left (\frac{\overline{Q}^2 + m^2}{m^2}\right ) + \frac{m^2}{2\overline{Q}^2}\log\left (\frac{\overline{Q}^2 + m^2}{m^2}\right ),
\end{multline}
and
\begin{multline}
\label{eq:Lfunction}
L(\gamma;z) =  \mathrm{Li}_{2}\left (\frac{1}{1-\frac{1}{2z}(1-\gamma)} \right )  + \mathrm{Li}_{2}\left (\frac{1}{1-\frac{1}{2(1-z)}(1-\gamma)} \right )   \\
+ \mathrm{Li}_{2}\left (\frac{1}{1-\frac{1}{2z}(1+\gamma)} \right )  + \mathrm{Li}_{2}\left (\frac{1}{1-\frac{1}{2(1-z)}(1+\gamma)}  \right ),
\end{multline}
with
\begin{align}
    \gamma &=\sqrt{1+\frac{4 m^2}{Q^2}}, \\
    \overline{Q}^2 &=z(1-z)Q^2 .
\end{align}
We also note that in the massless quark limit ($\gamma \to 1$) this contribution simplifies to 
\begin{equation}
    L(\gamma=1,z) = \frac{\pi^2}{6} - \frac{1}{2} \ln^2 \frac{z}{1-z}.
\end{equation}

The contribution on the 3rd line of Eq~\eqref{eq:NLO_dip_L} includes a sum of two special functions
\begin{equation}
\label{eq:IVsumFT}
\widetilde{\mathcal{I}}_{\mathcal{V}}(z,\xt_{01}) =  \widetilde{\mathcal{I}}_{\mathcal{V}_{(a)+(b)}}(z,\xt_{01})  +  \widetilde{\mathcal{I}}_{\mathcal{V}_{(c)+(d)}}(z,\xt_{01}) 
\end{equation}
defined as
\begin{equation}
\label{eq:J1int}
\begin{split}
\widetilde{\mathcal{I}}_{\mathcal{V}_{(a)+(b)}}(z,\xt_{01})  =  \int_{0}^{1} \frac{\dd \xi}{\xi}\biggl [-&\frac{2\log(\xi)}{(1-\xi)} + \frac{(1+\xi)}{2}\biggr ]\Biggl \{2K_0\left (\vert \xt_{01}\vert\sqrt{\overline{Q}^2+m^2}\right ) \\
& - K_0\left (\vert \xt_{01}\vert\sqrt{\overline{Q}^2+m^2 + \frac{(1-z)\xi}{1-\xi}m^2}\right ) \\
&- K_0\left (\vert \xt_{01}\vert\sqrt{\overline{Q}^2+m^2 + \frac{z\xi}{1-\xi}m^2}\right )\Biggr \},
\end{split}
\end{equation}
and 
\begin{equation}
\label{eq:J2int}
\begin{split}
\widetilde{\mathcal{I}}_{\mathcal{V}_{(c)+(d)}}(z,\xt_{01})  = m^2\int_{0}^{1}\dd \xi\int_{0}^{1}\dd x \Biggl \{& \Biggl [K_0\left (\vert \xt_{01}\vert\sqrt{\overline{Q}^2+m^2}\right ) - K_0\left (\vert \xt_{01}\vert\sqrt{\frac{\overline{Q}^2+m^2}{1-x} + \kappa}\right )\Biggr ]\\
& \hspace{1.7cm} \times
\frac{C^{L}_{m}}{(1-\xi)(1-x)\biggl [x(1-\xi) + \frac{\xi}{(1-z)}\biggr ]\biggl [\frac{x\left (\overline{Q}^2 + m^2\right )}{(1-x)} + \kappa\biggr ]}\\
& \hspace{-0.3cm} + \Biggl [K_0\left (\vert \xt_{01}\vert\sqrt{\overline{Q}^2+m^2}\right ) - K_0\left (\vert \xt_{01}\vert\sqrt{\frac{\overline{Q}^2+m^2}{1-x} + \kappa'}\right )\Biggr ]\\
& \hspace{1.7cm} \times
\frac{\overline C^{L}_{m}}{(1-\xi)(1-x)\biggl [x(1-\xi) + \frac{\xi}{z}\biggr ]\biggl [\frac{x\left (\overline{Q}^2 + m^2\right )}{(1-x)} + \kappa'\biggr ]}
\Biggr \}.
\end{split}
\end{equation}
These contributions are included in the functions \begin{cppcode}
double ILdip_massive_Iab(double Q2, double z1, double r, double mf, double xi) 
double ILdip_massive_Icd(double Q2, double z1, double r, double mf, double xi, double x)
\end{cppcode}
with overall prefactors included in the Cuba integrand 
\begin{cppcode}
int integrand_dip_massive(const int *ndim, const double x[], const int *ncomp, double *f, void *userdata)
\end{cppcode}
Here the integration variables are $|\xt_{01}|, z, \xi$ and, in the case of $\widetilde{\mathcal{I}}_{\mathcal{V}_{(a) + (b)}}$, also $x$. These integrals can be directly calculated by using adaptive Monte Carlo algorithms from Cuba.

The coefficients   $C^{L}_{m}$ and  $\overline C^{L}_{m}$ used above are 
\begin{equation}\label{eq:defClm}
C^{L}_{m} = \frac{z^2(1-\xi)}{(1-z)}\Biggl \{-\xi^2 + x(1-\xi)\frac{\biggl [1 + (1-\xi)\left (1 + \frac{z\xi}{(1-z)}\right )\biggr ]}{\biggl [x(1-\xi) + \frac{\xi}{(1-z)} \biggr ]} \Biggr \}.
\end{equation}
\begin{equation}\label{eq:defClmbar}
\overline C^{L}_{m} = \frac{(1-z)^2(1-\xi)}{z}\Biggl \{-\xi^2 + x(1-\xi)\frac{\biggl [1 + (1-\xi)\left (1 + \frac{(1-z)\xi}{z}\right )\biggr ]}{\biggl [x(1-\xi) + \frac{\xi}{z} \biggr ]} \Biggr \}.
\end{equation}
Furthermore  $\kappa$ and $\kappa'$ are defined as: 
\begin{equation}
\begin{split}
\kappa & = \frac{\xi m^2}{(1-\xi)(1-x)\biggl [x(1-\xi) + \frac{\xi}{(1-z)}\biggr ]}\biggl [\xi(1-x)+ x\left (1 - \frac{z(1-\xi)}{(1-z)}\right ) \biggr ],\\
\kappa' & = \frac{\xi m^2}{(1-\xi)(1-x)\biggl [x(1-\xi) + \frac{\xi}{z}\biggr ]}\biggl [\xi(1-x)+ x\left (1 - \frac{(1-z)(1-\xi)}{z}\right ) \biggr ].
\end{split}
\end{equation}

\subsection{Transverse photon, dipole contribution}
\label{appendix:special_fun_T_dip}
The special functions in Eq.~\eqref{eq:sigma_dip_T} read
\begin{equation}
\label{eq:fVfNfVMS}
\begin{split}
f_{\mathcal{V}} & = \biggl \{\frac{5}{2} - \frac{\pi^2}{3} + \log^2\left (\frac{z}{1-z} \right ) + \Omega^T_{\mathcal{V}} + L\biggr \} \kaz K_{1}\left (\vert \xt_{01}\vert \kaz \right )  + \widetilde{I}_{{V}}^T, \\
f_{\mathcal{N}} &  =  \Omega_{\mathcal{N}}^T \,  \kaz  K_{1}\left (\vert \xt_{01}\vert \kaz \right ) + \widetilde{I}^T_{{N}},  \\
f_{{VMS}}  & = \Biggl \{ 3 - \frac{\pi^2}{3}  + \log^2 \left (\frac{z}{1-z} \right )  + \Omega_{\mathcal{V}}^T + L  \Biggr \}K_{0}\left (\vert \xt_{01}\vert \kaz \right )   + \widetilde{I}_{{VMS}}.
\end{split}
\end{equation}
Here $L=L(\gamma;z)$ is defined in Eq.~\eqref{eq:Lfunction}.
Recall that $\kaz = \sqrt{z(1-z)Q^2 + m^2}$. The terms not involving any numerical integrals read
\begin{multline}
\label{eq:SigmaOmega}
\Omega_{\mathcal{V}}^T  = \left (1 + \frac{1}{2z} \right )\biggl [\log(1-z) + \gamma \log \left (\frac{1 + \gamma}{1 + \gamma - 2z} \right ) \biggr ] - \frac{1}{2z}\biggl [ \left (z + \frac{1}{2} \right )(1-\gamma) +  \frac{m^2}{Q^2}\biggr ]\log \left (\frac{\overline{Q}^2 + m^2}{m^2} \right ) \\
+ \Bigg[ z \leftrightarrow 1-z \Bigg]
\end{multline}
and
\begin{multline}
\label{eq:degOmegaN}
\Omega_{\mathcal{N}}^T =   \frac{1+z-2z^2}{z}\Biggl [\log(1-z) + \gamma \log \left (\frac{1+\gamma}{1+\gamma-2z}  \right ) \Biggr ] + \frac{(1-z)}{z} \Biggl [ \biggl (\frac{1}{2} + z \biggr )(\gamma -1) - \frac{m^2}{Q^2} \Biggr ]\log \left (\frac{\overline{Q}^2 + m^2}{m^2} \right ) \\ 
- \Bigg[ z \leftrightarrow 1-z \Bigg]
\end{multline}
These are implemented in
\begin{cppcode}
double OmegaT_V(double Q, double z, double mf)
double OmegaT_N(double Q, double z, double mf) 
\end{cppcode}

In Eq.~\eqref{eq:fVfNfVMS} we also defined
\begin{equation}
    \label{eq:IVT}
    \widetilde{I}_{V}^T =  \widetilde{I}_{{V},1}^T + \widetilde{I}_{{V,2}}^T
\end{equation}
where
  \begin{equation}
\label{eq:IVT1}
\begin{split}
\widetilde{I}_{{V},1}^T &=
  \int_0^1 \dd \xi \Bigg[ \frac{1}{\xi} \left (\frac{2\log(\xi)}{1-\xi} - \frac{(1+\xi)}{2} \right ) \Biggl \{\sqrt{\kaz^2 + \frac{\xi(1-z)}{(1-\xi)}m^2}\,K_{1}\left (\vert \xt_{01}\vert \sqrt{\kaz^2 + \frac{\xi(1-z)}{(1-\xi)}m^2}\right ) - [\xi \to 0]\Biggr \}\\
 & - \left (\frac{\log(\xi)}{(1-\xi)^2} + \frac{z}{(1-\xi)} + \frac{z}{2} \right ) \frac{(1-z)m^2}{\sqrt{\kaz^2 + \frac{\xi(1-z)}{(1-\xi)}m^2}}K_{1}\left (\vert \xt_{01}\vert \sqrt{\kaz^2 + \frac{\xi(1-z)}{(1-\xi)}m^2}\right ) \Bigg] + \Bigg[ z \leftrightarrow 1-z \Bigg].
 \end{split}
 \end{equation}
 In the code this is implemented in
 \begin{cppcode}
double IT_V1(double Q, double z, double mf, double r, double xi)
 \end{cppcode}

 The term $\widetilde{I}_{{V,2}}^T$ has two numerical integrals and reads
 \begin{equation}
     \begin{split}
     \label{eq:IVT2}
\widetilde{I}_{{V},2}^T &=
  -\int_{0}^{z} \dd \chi \int_{0}^{\infty} \dd u \Bigg\{\frac{1}{(1-\chi)}
\frac{1 }{u(u + 1)} \frac{ m^2}{\kac^2} \biggl [2\chi + \left (\frac{u}{1+u}\right )^2 \frac{1}{z}(z-\chi)(1-2\chi) \biggr ]\\
& \hspace{2cm} \times \Biggl \{\sqrt{\kaz^2 + u\frac{(1-z)}{(1-\chi)}\kac^2}K_{1}\left (\vert \xt_{01}\vert \sqrt{\kaz^2 + u\frac{(1-z)}{(1-\chi)}\kac^2}\right ) - [u \to 0] \Biggr \}\\
& + \frac{1}{(1-\chi)^2}
\frac{1}{(u + 1)} \, \left (z - \chi\right )\left [1 - \frac{2u}{1+u}\left (z - \chi\right ) +  \left (\frac{u}{1+u} \right )^2 \frac{1}{z}\left (z - \chi\right )^2 \right ]\\
& \hspace{2cm}\times \frac{m^2}{\sqrt{\kaz^2 + u\frac{(1-z)}{(1-\chi)}\kac^2}}K_{1}\left (\vert \xt_{01}\vert \sqrt{\kaz^2 + u\frac{(1-z)}{(1-\chi)}\kac^2}\right ) \Bigg\}\\
& +  \biggl [z \leftrightarrow 1-z \biggr ].
\end{split}
\end{equation}
The integrand of this function is implemented in 
\begin{cppcode}
double IT_V2(double Q, double z, double mf, double r, double y_chi, double y_u)
\end{cppcode}
As the Cuba integration routines integrate over the unit hypercube, when computing $\widetilde{I}_{{V},2}^T$ we first make a change of variables 
\begin{equation}
\label{eq:change_of_variables_2dint}
\begin{split}
    \chi &= z y_\chi \\
    u &= \frac{1-y_u}{y_u}
\end{split}
\end{equation}
as $y_\chi$ (parameter \code{y_chi}) and $y_u$ (parameter \code{y_u}) are integrated over $[0,1]$.
The same change of variables is  performed when evaluating the contribution $\widetilde{I}^T_{{N}}$ in Eq.~\eqref{eq:fVfNfVMS}, where one also has two numerical integrals 
\begin{equation}
\label{eq:INT}
\begin{split}
  \widetilde{I}^T_{{N}}   & = \frac{2(1-z)}{z} \int_0^z \dd \chi \int_0^\infty \frac{\dd u}{(u+1)^3} \Biggl \{\left [(2+u)u z + u^2\chi\right ]\sqrt{\kaz^2 + u \frac{(1-z)}{(1-\chi)}\kac^2} \, K_1\left (\vert \xt_{01}\vert  \sqrt{\kaz^2 + u \frac{(1-z)}{(1-\chi)}\kac^2} \right ) 
  \\
&
+ \frac{m^2}{\kac^2} 
\biggl (\frac{z}{1-z} + \frac{\chi}{1 - \chi} [u  - 2z - 2u\chi] \biggr )
\Biggl [\sqrt{\kaz^2 + u \frac{(1-z)}{(1-\chi)}\kac^2} \, K_1\left (\vert \xt_{01}\vert  \sqrt{\kaz^2 + u \frac{(1-z)}{(1-\chi)}\kac^2} \right ) -[u\to 0] \Biggr ]\Biggr \}\\
& -  \biggl [z \leftrightarrow 1-z \biggr ].
\end{split}
\end{equation}
This integrand is implemented in
\begin{cppcode}
double IT_N(double Q, double z, double mf, double r, double y_chi, double y_u)
\end{cppcode}

Furthermore we define
\begin{equation}
    \widetilde{I}_{VMS}  = \widetilde{I}_{VMS,1} + \widetilde{I}_{VMS,2}  
\end{equation}
where
\begin{equation}
\label{eq:IVMST1}
\begin{split}
\widetilde{I}_{VMS,1}  & = \int_{0}^{1} \dd \xi \Bigg[ \frac{1}{\xi} \left (\frac{2\log(\xi)}{(1-\xi)} - \frac{(1+\xi)}{2}\right )\Biggl \{ K_{0}\left (\vert \xt_{01}\vert \sqrt{\kaz^2  + \frac{\xi(1-z)}{(1-\xi)}m^2}\right )  -[\xi \to 0] \Biggr \}\\
& + \left (-\frac{3(1-z)}{2(1-\xi)} + \frac{(1-z)}{2}\right ) K_{0}\left (\vert \xt_{01}\vert \sqrt{\kaz^2  + \frac{\xi(1-z)}{(1-\xi)}m^2}\right ) \Bigg] + \Bigg[z \leftrightarrow 1-z\Bigg]
\end{split}
\end{equation}
and
\begin{equation}
\label{eq:IVMST2}
    \begin{split}
\widetilde{I}_{VMS,2}
& = \int_{0}^{z} \frac{\dd \chi}{(1-\chi)}\int_{0}^{\infty} \frac{\dd u}{(u+1)^2} \biggl \{ -z - \frac{u}{(1+u)}\frac{(z +u\chi)}{z}(\chi  - (1-z)) \biggr \} 
K_{0}\left (\vert \xt_{01}\vert \sqrt{\kaz^2  + u\frac{(1-z)}{(1-\chi)}\kac^2}\right )\\
& + \int_{0}^{z} \dd \chi \int_{0}^{\infty} \frac{\dd u}{(u+1)^3}  \biggl \{\frac{\kaz^2}{\kac^2} \biggl [1 + u\frac{\chi(1-\chi)}{z(1-z)}\biggr ] - \frac{m^2}{\kac^2} \frac{\chi}{(1-\chi)}\bigg [2\frac{(1+u)^2}{u} +  \frac{u}{z(1-z)}\left (z - \chi\right )^2 \biggr ]\biggr \}\\
&  \hspace{2cm}  \times \Biggl \{K_{0}\left (\vert \xt_{01}\vert \sqrt{\kaz^2  + u\frac{(1-z)}{(1-\chi)}\kac^2}\right ) - [u \to 0]    \Biggr \}\\
& +  \biggl [z \leftrightarrow 1-z \biggr ].
\end{split}
\end{equation}
These integrands are implemented in
\begin{cppcode}
double IT_VMS1(double Q, double z, double mf, double r, double xi)
double IT_VMS2(double Q, double z, double mf, double r, double y_chi, double y_u) 
\end{cppcode}
Again, when evaluating $\widetilde{I}_{VMS,2}$ with two numerical integrals, we first perform the change of variables~\eqref{eq:change_of_variables_2dint}.

\subsection{Longitudinal photon, \texorpdfstring{$q \bar q g$}{qqg} contribution}
\label{sec:appendix_special_fun_qqg_L}

The special functions that appear in Eq.~\eqref{eq:L_qqg_decomposition} are
\begin{align}
\label{eq:qqg_L_I1}
I_1^L= \frac{1}{z_2} \Bigg\{ 
& \frac{1}{|\mathbf{x}_{20}|^2} z_1^2 \left[ 2 z_0 (z_0 + z_2) + z_2^2 \right]
\Bigg[
K_0\left(
\sqrt{\overline{Q}_{(k)}^{\,2} + m^2}
\sqrt{|\mathbf{x}_{3;(k)}|^2 + \omega_{(k)} |\mathbf{x}_{20}|^2}
\right)^{2}
\Bigg]
\nonumber \\
&+
\frac{1}{|\mathbf{x}_{21}|^2} z_0^2
\left[ 2 z_1 (z_1 + z_2) + z_2^2 \right]
\Bigg[
K_0\left(
\sqrt{\overline{Q}_{(l)}^{\,2} + m^2}
\sqrt{|\mathbf{x}_{3;(l)}|^2 + \omega_{(l)} |\mathbf{x}_{21}|^2}
\right)^{2}
\Bigg]
\nonumber \\
&-
2 z_1 z_0
\left[
z_1 (1 - z_1) + z_0 (1 - z_0)
\right]
\frac{\mathbf{x}_{20} \cdot \mathbf{x}_{21}}
{|\mathbf{x}_{20}|^2 |\mathbf{x}_{21}|^2}
\nonumber \\
&\times
K_0\left(
\sqrt{\overline{Q}_{(k)}^{\,2} + m^2}
\sqrt{|\mathbf{x}_{3;(k)}|^2 + \omega_{(k)} |\mathbf{x}_{20}|^2}
\right)
K_0\left(
\sqrt{\overline{Q}_{(l)}^{\,2} + m^2}
\sqrt{|\mathbf{x}_{3;(l)}|^2 + \omega_{(l)} |\mathbf{x}_{21}|^2}
\right)
\Bigg\}
\end{align}
In the code this is implemented in
\begin{cppcode}
double ILNLOqg_massive_tripole_part_I1(double Q2, double mf, double z1, double z2, double x01sq, double x02sq, double x21sq)
\end{cppcode}
and is multiplied by $N_{012}$.

The UV subtraction term proportional to $N_{01}$ reads
\begin{align}
\label{eq:qqg_L_uv}
I_\mathrm{UV}^L = \frac{1}{z_2} \Bigg\{ 
& \frac{1}{|\mathbf{x}_{20}|^2} z_1^2 \left[ 2 z_0 (z_0 + z_2) + z_2^2 \right]
\Bigg[
- \frac{
e^{-|\mathbf{x}_{20}|^2/(|\mathbf{x}_{01}|^2 e^{\gamma_E})}
}{
|\mathbf{x}_{20}|^2
}
K_0\left(
|\mathbf{x}_{01}|
\sqrt{\overline{Q}_{(k)}^{\,2} + m^2}
\right)^{2}
\Bigg]
\nonumber \\
&+
\frac{1}{|\mathbf{x}_{21}|^2} z_0^2
\left[ 2 z_1 (z_1 + z_2) + z_2^2 \right]
\Bigg[
- \frac{
e^{-|\mathbf{x}_{21}|^2/(|\mathbf{x}_{01}|^2 e^{\gamma_E})}
}{
|\mathbf{x}_{21}|^2
}
K_0\left(
|\mathbf{x}_{01}|
\sqrt{\overline{Q}_{(l)}^{\,2} + m^2}
\right)^{2}
\Bigg].
\end{align}
This is implemented in 
\begin{cppcode}
double ILNLOqg_massive_dipole_uvsub(double Q2, double mf, double z1, double z2, double x01sq, double x02sq, double x21sq)
\end{cppcode}

The remaining contributions with one or two additional numerical integrations read
\begin{equation}
\label{eq:IL2}
\begin{split}
    I^L_2 =& \int_0^1\dd{y} \frac{1}{z_2}\left\{ z_1^2 \left[2z_0 (z_0+z_2)+z_2^2 \right] \frac{1}{4}  \overline g^{(1;2)}_{(\textrm{k})}(y) K_0\left( \sqrt{\overline Q^2_{(\textrm{k})}+m^2} \sqrt{ |\xt_{3;(\textrm{k})}|^2 + \omega_{(\textrm{k})} |\xt_{20)}|^2}  \right)  \right. \\
    &\left.+ z_0^2 \left[2z_1 (z_1+z_2)+z_2^2 \right] \frac{1}{4}   \overline g^{(1;2)}_{(\textrm{l})}(y) K_0\left( \sqrt{\overline Q^2_{(\textrm{l})}+m^2} \sqrt{ |\xt_{3;(\textrm{l})}|^2 + \omega_{(\textrm{l})} |\xt_{21)}|^2}  \right)  \right. \\
    & -\frac{1}{4}  z_1 z_0 \left[z_1 (1-z_1) + z_0 (1-z_0) \right] \left(\xt_{20} \vdot \xt_{21}\right)  \\
    &\times \left. \left[
    \frac{1}{|\xt_{21}|^2 }\overline g^{(1;2)}_{(\textrm{k})}(y) K_0\left( \sqrt{\overline Q^2_{(\textrm{l})}+m^2} \sqrt{ |\xt_{3;(\textrm{l})}|^2 + \omega_{(\textrm{l})} |\xt_{21)}|^2}  \right) \right.\right. \\
    & \left.\left. +
    \frac{1}{|\xt_{20}|^2 }  \overline g^{(1;2)}_{(\textrm{l})}(y) K_0\left( \sqrt{\overline Q^2_{(\textrm{k})}+m^2} \sqrt{ |\xt_{3;(\textrm{k})}|^2 + \omega_{(\textrm{k})} |\xt_{20)}|^2}  \right) \right]
    \right\}
\end{split}
\end{equation}
and
\begin{equation}
\label{eq:IL3}
\begin{split}
    I_3^L
    =& \int_0^1 \dd{y_{1}}\int_0^1 \dd{y_{2}}
    \frac{1}{z_2}
    \left\{ z_1^2 \left[2z_0 (z_0+z_2)+z_2^2 \right] \frac{|\xt_{20}|^2}{64}  \overline g^{(1;2)}_{(\textrm{k})}(y_1)\overline g^{(1;2)}_{(\textrm{k})}(y_2) \right. \\
    & +  z_0^2 \left[2z_1 (z_1+z_2)+z_2^2 \right] \frac{|\xt_{21}|^2}{64}  \overline g^{(1;2)}_{(\textrm{l})}(y_1)\overline g^{(1;2)}_{(\textrm{l})}(y_2)\\
    &-\frac{1}{32}  z_1 z_0 \left[z_1 (1-z_1) + z_0 (1-z_0) \right] \left(\xt_{20} \vdot \xt_{21}\right)
     \overline g^{(1;2)}_{(\textrm{k})}(y_1) \overline g^{(1;2)}_{(\textrm{l})}(y_2)  \\
    &\left.+ \frac{m^2}{16} z_2^4  \left[ \left(\frac{z_1}{z_0+z_2}\right)^2 g^{(1;1)}_{(\textrm{k})}(y_1) g^{(1;1)}_{(\textrm{k})}(y_2)   + \left(\frac{z_0}{z_1+z_2}\right)^2 g^{(1;1)}_{(\textrm{l})}(y_1) g^{(1;1)}_{(\textrm{l})}(y_2)\right. \right.\\
    &\left. \left.
    - 2  \frac{z_0}{z_1+z_2} \frac{z_1}{z_0+z_2}  g^{(1;1)}_{(\textrm{k})}(y_1)  g^{(1;1)}_{(\textrm{l})}(y_2) \right] \right\} .
\end{split}
\end{equation}
These integrands are implemented in
\begin{cppcode}
double ILNLOqg_massive_tripole_part_I2(double Q2, double mf, double z1, double z2, double x01sq, double x02sq, double x21sq, double y_t)
double ILNLOqg_massive_tripole_part_I3(double Q2, double mf, double z1, double z2, double x01sq, double x02sq, double x21sq, double y_t1, double y_t2) 
\end{cppcode}

\subsection{Transverse photon, \texorpdfstring{$q \bar q g$}{qqg} contribution}
\label{sec:appendix_special_fun_qqg_T}

The UV subtraction term proportional to $N_{01}$ in Eq.~\eqref{eq:qqg_T_decomposition} is
\begin{multline}
    \label{eq:qqg_T_uv}
    I^{T}_{UV}
= \frac{1}{z_2}
\Bigg\{
\frac{2 z_0 (z_0 + z_2) + z_2^2}{(z_0 + z_2)^2}
\left[ 1 - 2 z_1 (1 - z_1) \right]\frac{\overline{Q}_{(j)}^{\,2} + m^2}{|\xt_{2;(j)}|^2}
\\
 \times
\left[
- e^{-|\xt_{2;(j)}|^2 / (|\xt_{01}|^2 e^{\gamma_E})}
\left[
K_1\left(
|\xt_{01}|\sqrt{\overline{Q}_{(j)}^{\,2} + m^2}
\right)
\right]^2
\right]
\\
 + \frac{2 z_1 (z_1 + z_2) + z_2^2}{(z_1 + z_2)^2}
\left[ 1 - 2 z_0 (1 - z_0) \right] \frac{\overline{Q}_{(k)}^{\,2} + m^2}{|\xt_{2;(k)}|^2}
\\
\times
\left[
- e^{-|\xt_{2;(k)}|^2 / (|\xt_{01}|^2 e^{\gamma_E})}
\left[
K_1\left(
|\xt_{01}|\sqrt{\overline{Q}_{(k)}^{\,2} + m^2}
\right)
\right]^2
\right]
\Bigg\} \\
+
\frac{m^2}{z_2}
\Bigg\{
-\frac{2z_0(z_0+z_2)+z_2^2}{(z_0 + z_2)^2}
\, e^{-|\mathbf{x}_{2;(j)}|^2 / (|\mathbf{x}_{01}|^2 e^{\gamma_E})}
\left[
K_0\left(
|\mathbf{x}_{01}| \sqrt{\overline{Q}_{(j)}^{\,2} + m^2}
\right)
\right]^2
\\
-\frac{2z_1(z_1+z_2)+z_2^2}{(z_1 + z_2)^2}
\, e^{-|\mathbf{x}_{2;(k)}|^2 / (|\mathbf{x}_{01}|^2 e^{\gamma_E})}
\left[
K_0\left(
|\mathbf{x}_{01}| \sqrt{\overline{Q}_{(k)}^{\,2} + m^2}
\right)
\right]^2
\Bigg\}.
\end{multline}
This is implemented in
\begin{cppcode}
double ITNLOqg_massive_dipole_uvsub(double Q2, double mf, double z1, double z2, double x01sq, double x02sq, double x21sq)
\end{cppcode}
The term in Eq.~\eqref{eq:qqg_T_decomposition} proportional to $N_{012}$ with no additional numerical integrations is
\begin{multline}
\label{eq:qqg_T_1}
    I_1^T = \frac{1}{z_2}
\Bigg\{
\frac{1}{(z_0 + z_2)^2}
\left[2 z_0 (z_0 + z_2) + z_2^2\right]
\left[1 - 2 z_1 (1 - z_1)\right]
\,\frac{\overline{Q}_{(j)}^{\,2} + m^2}{|\xt_{2;(j)}|^2}
\frac{|\xt_{3;(j)}|^2}{|\xt_{3;(j)}|^2 + \omega_{(j)}|\xt_{2;(j)}|^2} \\
\times 
\left[
K_1\left(
\sqrt{|\xt_{3;(j)}|^2 + \omega_{(j)} |\xt_{2;(j)}|^2}
\sqrt{\overline{Q}_{(j)}^{\,2} + m^2}
\right)
\right]^2
\\
+
\frac{1}{(z_1 + z_2)^2}
\left[2 z_1 (z_1 + z_2) + z_2^2\right]
\left[1 - 2 z_0 (1 - z_0)\right]
\,\frac{\overline{Q}_{(k)}^{\,2} + m^2}{|\xt_{2;(k)}|^2}
\frac{|\xt_{3;(k)}|^2}{|\xt_{3;(k)}|^2 + \omega_{(k)}|\xt_{2;(k)}|^2} \\
\times 
\left[
K_1\left(
\sqrt{|\xt_{3;(k)}|^2 + \omega_{(k)} |\xt_{2;(k)}|^2}
\sqrt{\overline{Q}_{(k)}^{\,2} + m^2}
\right)
\right]^2
\Bigg\} \\
+\frac{1}{z_2} \, m^2
\Bigg\{
\frac{1}{(z_0 + z_2)^2}
\left[ 2 z_0 (z_0 + z_2) + z_2^2 \right]
\frac{1}{|\mathbf{x}_{2;(j)}|^2}
\left[
K_0\!\left(
\sqrt{\overline Q_{(j)}^{2} + m^{2}}\,
\sqrt{|\mathbf{x}_{3;(j)}|^2 + \omega_{(j)} |\mathbf{x}_{2;(j)}|^2}
\right)
\right]^2
\\
+
\frac{1}{(z_1 + z_2)^2}
\left[ 2 z_1 (z_1 + z_2) + z_2^2 \right]
\frac{1}{|\mathbf{x}_{2;(k)}|^2}
\left[
K_0\!\left(
\sqrt{\overline Q_{(k)}^{2} + m^{2}}\,
\sqrt{|\mathbf{x}_{3;(k)}|^2 + \omega_{(k)} |\mathbf{x}_{2;(k)}|^2}
\right)
\right]^2
\Bigg\}
\end{multline}
This is implemented in
\begin{cppcode}
double IT_tripole_jk_I1(double Q, double mf, double z1, double z2, double x01sq, double x02sq, double x21sq)
double IT_tripole_jkm_I1(double Q, double mf, double z1, double z2, double x01sq, double x02sq, double x21sq) 
\end{cppcode}
The second function computes the part directly proportional to $m^2$.

The two other terms with the same dimension for the numerical integral and the same Wilson line structure (proportional to $N_{012}$) read
\begin{multline}
\label{eq:qqg_T_1F}
I^{T}_{1F}
=
\frac{1}{2z_2}\,
\Bigg\{
\frac{4}{(z_0+z_2) (z_1+z_2)}
\Big[
z_2 (z_0-z_1)^2
\big[
(\mathbf{x}_{3;(j)}\cdot\mathbf{x}_{2;(j)})
(\mathbf{x}_{3;(k)}\cdot\mathbf{x}_{2;(k)})
-
(\mathbf{x}_{3;(j)}\cdot\mathbf{x}_{2;(k)})
(\mathbf{x}_{3;(k)}\cdot\mathbf{x}_{2;(j)})
\big]
\\
\quad
-\,[z_1(z_0+z_2)+z_0(z_1+z_2)]
\big[
z_0(z_0+z_2)+z_1(z_1+z_2)
\big]
(\mathbf{x}_{2;(j)}\cdot\mathbf{x}_{3;(j)})
(\mathbf{x}_{2;(k)}\cdot\mathbf{x}_{3;(k)})
\Big]
\\[8pt]
\times
\frac{1}{|\mathbf{x}_{2;(j)}|^2}\,
\sqrt{\frac{\overline{Q}_{(j)}^{\,2}+m^2}{|\mathbf{x}_{3;(j)}|^2+\omega_{(j)}|\mathbf{x}_{2;(j)}|^2}}\,
K_1\left(
\sqrt{\overline{Q}_{(j)}^{\,2}+m^2}
\sqrt{|\mathbf{x}_{3;(j)}|^2+\omega_{(j)}|\mathbf{x}_{2;(j)}|^2}
\right)
\\
\times
\frac{1}{|\mathbf{x}_{2;(k)}|^2}\,
\sqrt{\frac{\overline{Q}_{(k)}^{\,2}+m^2}{|\mathbf{x}_{3;(k)}|^2+\omega_{(k)}|\mathbf{x}_{2;(k)}|^2}}\,
K_1\left(
\sqrt{\overline{Q}_{(k)}^{\,2}+m^2}
\sqrt{|\mathbf{x}_{3;(k)}|^2+\omega_{(k)}|\mathbf{x}_{2;(k)}|^2}
\right)
\\
-\frac{(z_0+z_2)z_1z_2}{(z_1+z_2)^2}\,
(\mathbf{x}_{2;(j)}\cdot\mathbf{x}_{3;(j)})\,\mathcal{H}_{(k)}
\,
\frac{1}{|\mathbf{x}_{2;(j)}|^2}\,
\sqrt{
\frac{\overline{Q}_{(j)}^{\,2}+m^2}{|\mathbf{x}_{3;(j)}|^2+\omega_{(j)}|\mathbf{x}_{2;(j)}|^2}}
K_1\left(
\sqrt{\overline{Q}_{(j)}^{\,2}+m^2}
\sqrt{|\mathbf{x}_{3;(j)}|^2+\omega_{(j)}|\mathbf{x}_{2;(j)}|^2}
\right)
\\
+\frac{(z_1+z_2)z_0z_2}{(z_0+z_2)^2}\,
(\mathbf{x}_{2;(k)}\cdot\mathbf{x}_{3;(k)})\,\mathcal{H}_{(j)}
\,
\frac{1}{|\mathbf{x}_{2;(k)}|^2}\,
\sqrt{
\frac{\overline{Q}_{(k)}^{\,2}+m^2}{|\mathbf{x}_{3;(k)}|^2+\omega_{(k)}|\mathbf{x}_{2;(k)}|^2}}
K_1\left(
\sqrt{\overline{Q}_{(k)}^{\,2}+m^2}
\sqrt{|\mathbf{x}_{3;(k)}|^2+\omega_{(k)}|\mathbf{x}_{2;(k)}|^2}
\right)
\\
-\frac{z_0^2 z_1 z_2}{(z_0+z_2)^3}\,
(\mathbf{x}_{2;(j)}\cdot\mathbf{x}_{3;(j)})\,\mathcal{H}_{(j)}
\,
\frac{1}{|\mathbf{x}_{2;(j)}|^2}\,
\sqrt{
\frac{\overline{Q}_{(j)}^{\,2}+m^2}{|\mathbf{x}_{3;(j)}|^2+\omega_{(j)}|\mathbf{x}_{2;(j)}|^2}}
K_1\left(
\sqrt{\overline{Q}_{(j)}^{\,2}+m^2}
\sqrt{|\mathbf{x}_{3;(j)}|^2+\omega_{(j)}|\mathbf{x}_{2;(j)}|^2}
\right)
\\
+\frac{z_1^2 z_0 z_2}{(z_1+z_2)^3}\,
(\mathbf{x}_{2;(k)}\cdot\mathbf{x}_{3;(k)})\,\mathcal{H}_{(k)}
\,
\frac{1}{|\mathbf{x}_{2;(k)}|^2}\,
\sqrt{
\frac{\overline{Q}_{(k)}^{\,2}+m^2}{|\mathbf{x}_{3;(k)}|^2+\omega_{(k)}|\mathbf{x}_{2;(k)}|^2}}
K_1\left(
\sqrt{\overline{Q}_{(k)}^{\,2}+m^2}
\sqrt{|\mathbf{x}_{3;(k)}|^2+\omega_{(k)}|\mathbf{x}_{2;(k)}|^2}
\right)
\\[12pt]
+\frac{z_2^2 z_0^2 }{8(z_0+z_2)^4}\,[\mathcal{H}_{(j)}]^2
+\frac{z_2^2 z_1^2}{8(z_1+z_2)^4}\,[\mathcal{H}_{(k)}]^2
\Bigg\}
\end{multline}
and
\begin{multline}
\label{eq:qqg_T_1Fm}
    I_{1F_m}^T= 
\frac{m^2}{2z_2}
\times
\Bigg\{
-\frac{1}{32 (z_0+z_2)(z_1+z_2)}
\left[
(2z_0 + z_2)(2z_1 + z_2) + z_2^{2}
\right]
\left( \mathbf{x}_{2;(j)} \cdot \mathbf{x}_{2;(k)} \right)
\\
\times
\frac{8}{|\mathbf{x}_{2;(j)}|^{2}}
K_{0} \left(
\sqrt{\overline{Q}_{(j)}^{2}+m^{2}}\,
\sqrt{\,|\mathbf{x}_{3;(j)}|^{2} + \omega_{(j)} |\mathbf{x}_{2;(j)}|^{2}}
\right)
\;
\frac{8}{|\mathbf{x}_{2;(k)}|^{2}}
K_{0} \left(
\sqrt{\overline{Q}_{(k)}^{2}+m^{2}}\,
\sqrt{\,|\mathbf{x}_{3;(k)}|^{2} + \omega_{(k)} |\mathbf{x}_{2;(k)}|^{2}}
\right)
\Bigg\}.
\end{multline}
Here we defined
\begin{multline}
\mathcal{H}_{(x)}=
    \int_{0}^{\infty} \frac{\dd u}{u^{2}}
\exp \left\{
-u\left[ \overline{Q}_{(x)}^{2} + m^{2}\,(1+\lambda_{(x)}) \right]
\right\}
\exp \left\{
-\frac{|\mathbf{x}_{3;(x)}|^{2} + \omega_{(x)} |\mathbf{x}_{2;(x)}|^{2}}{4u}
\right\} \\
=
4
\sqrt{
\frac{
\overline Q_{(x)}^{2} + m^{2}(1+\lambda_{(x)})
}{
|\mathbf{x}_{3;(x)}|^{2} + \omega_{(x)} |\mathbf{x}_{2;(x)}|^{2}
}
}
\;
K_{1} \left(
\sqrt{\overline{Q}_{(x)}^{2} + m^{2}(1+\lambda_{(x)})}
\;
\sqrt{|\mathbf{x}_{3;(x)}|^{2} + \omega_{(x)} |\mathbf{x}_{2;(x)}|^{2}}
\right).
\end{multline}
In the code these two integrands are implemented in
\begin{cppcode}
double IT_tripole_F_I1(double Q, double mf, double z1, double z2, double x01sq, double x02sq, double x21sq)
double IT_tripole_Fm_I1(double Q, double mf, double z1, double z2, double x01sq, double x02sq, double x21sq)
\end{cppcode}

The contributions with one additional numerical integral, with the Wilson line structure $N_{012}$, are
\begin{equation}
\label{eq:qqg_T_2}
    \begin{split}
    I^T_2=&  \int_0^1 \dd{y}  
    \frac{1}{z_2}
    \Bigg\{
\frac{1}{(z_0+z_2)^2} \left[ 2z_0(z_0+z_2)+z_2^2 \right] \left[ 1-2 z_1 (1-z_1) \right] \\
&\times  \overline g^{(2;2)}_{(\textrm{j})}(y) \frac{|\xt_{3;(\textrm{j})}|^2}{8} \sqrt{\frac{\overline Q^2_{(\textrm{j})}+m^2}{|\xt_{3;(\textrm{j})}|^2 + \omega_{(\textrm{j})} |\xt_{2;(\textrm{j})}|^2}} K_1\left( \sqrt{\overline Q^2_{(\textrm{j})}+m^2} \sqrt{ |\xt_{3;(\textrm{j})}|^2 + \omega_{(\textrm{j})} |\xt_{2;(\textrm{j})}|^2}  \right)  \\
&+
\frac{1}{(z_1+z_2)^2} \left[ 2z_1(z_1+z_2)+z_2^2 \right] \left[1-2z_0(1-z_0) \right] \\
& \times  \overline g_{(\textrm{k})}^{(2;2)}(y) \frac{|\xt_{3;(\textrm{k})}|^2}{8} \sqrt{\frac{\overline Q^2_{(\textrm{k})}+m^2}{|\xt_{3;(\textrm{k})}|^2 + \omega_{(\textrm{k})} |\xt_{2;(\textrm{k})}|^2}} K_1\left( \sqrt{\overline Q^2_{(\textrm{k})}+m^2} \sqrt{ |\xt_{3;(\textrm{k})}|^2 + \omega_{(\textrm{k})} |\xt_{2;(\textrm{k})}|^2}  \right)  
\Bigg\}
    \end{split}
\end{equation}

\begin{equation}
\label{eq:qqg_T_2m}
    \begin{split}
        I^T_{2m}  =\int_0^1 \dd{y} m^2 \frac{1}{z_2}
        &\Bigg\{
\frac{1}{(z_0+z_2)^2} \left[ 2z_0(z_0+z_2)+z_2^2 \right] 
\frac{ 1}{4} \overline g_{(\textrm{j})}^{(1;2)}( y) K_0\left( \sqrt{\overline Q^2_{(\textrm{j})}+m^2} \sqrt{ |\xt_{3;(\textrm{j})}|^2 + \omega_{(\textrm{j})} |\xt_{2;(\textrm{j})}|^2}  \right)  \\
+&
\frac{1}{(z_1+z_2)^2} \left[ 2z_1(z_1+z_2)+z_2^2 \right]
\frac{ 1}{4} \overline g_{(\textrm{k})}^{(1;2)}( y) K_0\left( \sqrt{\overline Q^2_{(\textrm{k})}+m^2} \sqrt{ |\xt_{3;(\textrm{k})}|^2 + \omega_{(\textrm{k})} |\xt_{2;(\textrm{k})}|^2}  \right) 
\Bigg\}
    \end{split}
\end{equation}

\begin{equation}
\label{eq:qqg_T_2F}
    \begin{split}
&I^T_{2F}  = 
\int_0^1 \dd{y} 
     \frac{1}{2} \frac{1}{z_2}  \\
&\times \Bigg\{
\frac{1}{4(z_0+z_2)(z_1+z_2)} 
\left\{ 
z_2 (z_0-z_1)^2 \left[ \left( \xt_{3;(\textrm{j})} \vdot \xt_{2;(\textrm{j})} \right)\left( \xt_{3;(\textrm{k})} \vdot \xt_{2;(\textrm{k})} \right)-\left( \xt_{3;(\textrm{j})} \vdot \xt_{2;(\textrm{k})} \right)\left( \xt_{3;(\textrm{k})} \vdot \xt_{2;(\textrm{j})} \right) \right] \right.\\
&-\left[z_1(z_0+z_2)+z_0(z_1+z_2) \right] \left[z_0(z_0+z_2)+z_1(z_1+z_2) \right] \left( \xt_{2;(\textrm{j})} \vdot \xt_{2;(\textrm{k})} \right)\left( \xt_{3;(\textrm{j})} \vdot \xt_{3;(\textrm{k})} \right)
\Bigg\} \\
&\times \bigg[
\overline g_{(\textrm{k})}^{(2;2)}(y) 
\frac{1}{|\xt_{2;(\textrm{j})}|^2} \sqrt{\frac{\overline Q^2_{(\textrm{j})}+m^2}{|\xt_{3;(\textrm{j})}|^2 + \omega_{(\textrm{j})} |\xt_{2;(\textrm{j})}|^2}} K_1\left( \sqrt{\overline Q^2_{(\textrm{j})}+m^2} \sqrt{ |\xt_{3;(\textrm{j})}|^2 + \omega_{(\textrm{j})} |\xt_{2;(\textrm{j})}|^2}  \right) \\
&+\overline  g_{(\textrm{j})}^{(2;2)}(y) 
\frac{1}{|\xt_{2;(\textrm{k})}|^2} \sqrt{\frac{\overline Q^2_{(\textrm{k})}+m^2}{|\xt_{3;(\textrm{k})}|^2 + \omega_{(\textrm{k})} |\xt_{2;(\textrm{k})}|^2}} K_1\left( \sqrt{\overline Q^2_{(\textrm{k})}+m^2} \sqrt{ |\xt_{3;(\textrm{k})}|^2 + \omega_{(\textrm{k})} |\xt_{2;(\textrm{k})}|^2}  \right)
\bigg]
\\
&-\frac{(z_0+z_2) z_1 z_2}{16(z_1+z_2)^2}  \left( \xt_{2;(\textrm{j})} \vdot \xt_{3;(\textrm{j})} \right)  \hcal_{(\textrm{k})}  \overline  g_{(\textrm{j})}^{(2;2)}(y) 
+\frac{(z_1+z_2)z_0 z_2}{16 (z_0+z_2)^2}  \left( \xt_{2;(\textrm{k})} \vdot \xt_{3;(\textrm{k})} \right)  \hcal_{(\textrm{j})} \overline g_{(\textrm{k})}^{(2;2)}(y)  \\
&-\frac{z_0^2 z_1 z_2}{16(z_0+z_2)^3}  \left( \xt_{2;(\textrm{j})} \vdot \xt_{3;(\textrm{j})} \right)  \hcal_{(\textrm{j})}  \overline  g_{(\textrm{j})}^{(2;2)}(y)
+\frac{z_1^2 z_0 z_2}{16 (z_1+z_2)^3}  \left( \xt_{2;(\textrm{k})} \vdot \xt_{3;(\textrm{k})} \right)  \hcal_{(\textrm{k})} \overline g_{(\textrm{k})}^{(2;2)}(y) 
\Bigg\}
    \end{split}
\end{equation}

\begin{equation}
\label{eq:qqg_T_2Fm}
    \begin{split}
&I^T_{2F_m}  = 
\int_0^1 \dd{y} 
        \frac{1}{z_2} \frac{m^2}{2} \\
&\times \Bigg\{
-\frac{z_0z_1z_2^2}{16(z_0+z_2)^3} \left( \xt_{3;(\textrm{j})} \vdot \xt_{2;(\textrm{j})} \right) 
 g_{(\textrm{j})}^{(2;1)}( y)  
 \frac{8}{|\xt_{2;(\textrm{j})}|^2} K_0\left( \sqrt{\overline Q^2_{(\textrm{j})}+m^2} \sqrt{ |\xt_{3;(\textrm{j})}|^2 + \omega_{(\textrm{j})} |\xt_{2;(\textrm{j})}|^2}  \right) \\
&+\frac{z_0z_1z_2^2}{16(z_1+z_2)^3} \left( \xt_{3;(\textrm{k})} \vdot \xt_{2;(\textrm{k})} \right) 
 g_{(\textrm{k})}^{(2;1)}( y) 
  \frac{8}{|\xt_{2;(\textrm{k})}|^2} K_0\left( \sqrt{\overline Q^2_{(\textrm{k})}+m^2} \sqrt{ |\xt_{3;(\textrm{k})}|^2 + \omega_{(\textrm{k})} |\xt_{2;(\textrm{k})}|^2}  \right) \\
&- \frac{1}{32(z_0+z_2)(z_1+z_2)} \left[(2z_0+z_2)(2z_1+z_2)+z_2^2 \right] \left( \xt_{2;(\textrm{j})} \vdot \xt_{2;(\textrm{k})} \right) \\
&\times \bigg\{
 \overline g_{(\textrm{k})}^{(1;2)}( y)
\frac{8}{|\xt_{2;(\textrm{j})}|^2} K_0\left( \sqrt{\overline Q^2_{(\textrm{j})}+m^2} \sqrt{ |\xt_{3;(\textrm{j})}|^2 + \omega_{(\textrm{j})} |\xt_{2;(\textrm{j})}|^2}  \right)\\
&+\overline g_{(\textrm{j})}^{(1;2)}( y)
\frac{8}{|\xt_{2;(\textrm{k})}|^2} K_0\left( \sqrt{\overline Q^2_{(\textrm{k})}+m^2} \sqrt{ |\xt_{3;(\textrm{k})}|^2 + \omega_{(\textrm{k})} |\xt_{2;(\textrm{k})}|^2}  \right)
\bigg\}\\
& -\frac{(z_0z_2)^2}{16(z_0+z_2)(z_1+z_2)^2} \left( \xt_{2;(\textrm{j})} \vdot \xt_{3;(\textrm{k})} \right)
 g_{(\textrm{k})}^{(2;1)}( y) 
 \times  \frac{8}{|\xt_{2;(\textrm{j})}|^2} K_0\left( \sqrt{\overline Q^2_{(\textrm{j})}+m^2} \sqrt{ |\xt_{3;(\textrm{j})}|^2 + \omega_{(\textrm{j})} |\xt_{2;(\textrm{j})}|^2}  \right) \\
&+\frac{(z_1z_2)^2}{16(z_0+z_2)^2(z_1+z_2)} \left( \xt_{3;(\textrm{j})} \vdot \xt_{2;(\textrm{k})} \right)
 g_{(\textrm{j})}^{(2;1)}( y)
 \times  \frac{8}{|\xt_{2;(\textrm{k})}|^2} K_0\left( \sqrt{\overline Q^2_{(\textrm{k})}+m^2} \sqrt{ |\xt_{3;(\textrm{k})}|^2 + \omega_{(\textrm{k})} |\xt_{2;(\textrm{k})}|^2}  \right) \\
& -\frac{z_0z_1 z_2^2}{16(z_0+z_2)^3}  \left( \xt_{2;(\textrm{j})} \vdot \xt_{3;(\textrm{j})} \right) 
 g_{(\textrm{j})}^{(1;1)}( y) \\
&\times\frac{16}{|\xt_{2;(\textrm{j})}|^2} \sqrt{\frac{\overline Q^2_{(\textrm{j})}+m^2}{|\xt_{3;(\textrm{j})}|^2 + \omega_{(\textrm{j})} |\xt_{2;(\textrm{j})}|^2}} K_1\left( \sqrt{\overline Q^2_{(\textrm{j})}+m^2} \sqrt{ |\xt_{3;(\textrm{j})}|^2 + \omega_{(\textrm{j})} |\xt_{2;(\textrm{j})}|^2}  \right) \\
&+\frac{z_0z_1 z_2^2}{16(z_1+z_2)^3}  \left( \xt_{2;(\textrm{k})} \vdot \xt_{3;(\textrm{k})} \right) 
g_{(\textrm{k})}^{(1;1)}( y) \\
&\times\frac{16}{|\xt_{2;(\textrm{k})}|^2} \sqrt{\frac{\overline Q^2_{(\textrm{k})}+m^2}{|\xt_{3;(\textrm{k})}|^2 + \omega_{(\textrm{k})} |\xt_{2;(\textrm{k})}|^2}} K_1\left( \sqrt{\overline Q^2_{(\textrm{k})}+m^2} \sqrt{ |\xt_{3;(\textrm{k})}|^2 + \omega_{(\textrm{k})} |\xt_{2;(\textrm{k})}|^2}  \right) \\
& - \frac{(z_0+z_2) z_2^2}{16(z_1+z_2)^2}   \left( \xt_{2;(\textrm{j})} \vdot \xt_{3;(\textrm{j})} \right)
g_{(\textrm{k})}^{(1;1)}( y) \\
&\times\frac{16}{|\xt_{2;(\textrm{j})}|^2} \sqrt{\frac{\overline Q^2_{(\textrm{k})}+m^2}{|\xt_{3;(\textrm{j})}|^2 + \omega_{(\textrm{j})} |\xt_{2;(\textrm{j})}|^2}} K_1\left( \sqrt{\overline Q^2_{(\textrm{j})}+m^2} \sqrt{ |\xt_{3;(\textrm{j})}|^2 + \omega_{(\textrm{j})} |\xt_{2;(\textrm{j})}|^2}  \right) \\
&+\frac{(z_1+z_2) z_2^2}{16(z_0+z_2)^2}   \left( \xt_{2;(\textrm{k})} \vdot \xt_{3;(\textrm{k})} \right) 
 g_{(\textrm{j})}^{(1;1)}( y) \\
&\times\frac{16}{|\xt_{2;(\textrm{k})}|^2} \sqrt{\frac{\overline Q^2_{(\textrm{k})}+m^2}{|\xt_{3;(\textrm{k})}|^2 + \omega_{(\textrm{k})} |\xt_{2;(\textrm{k})}|^2}} K_1\left( \sqrt{\overline Q^2_{(\textrm{k})}+m^2} \sqrt{ |\xt_{3;(\textrm{k})}|^2 + \omega_{(\textrm{k})} |\xt_{2;(\textrm{k})}|^2}  \right)\\
&+\frac{z_0 z_2^3}{4(z_0+z_2)^4} \hcal_{(\textrm{j})}
g_{(\textrm{j})}^{(1;1)}( y)  
+\frac{z_1 z_2^3}{4(z_1+z_2)^4} \hcal_{(\textrm{k})}
 g_{(\textrm{k})}^{(1;1)}( y) 
\Bigg\}
    \end{split}
\end{equation}

Similarly contributions with two additional numerical integrals, multiplied by the Wilson line correlator $N_{012}$, are
\begin{equation}
\label{eq:qqg_T_3}
    \begin{split}
    I^T_3  =&  \int_0^1 \dd{y_{1}} \int_0^1 \dd{y_{2}}  \frac{1}{z_2}
   \\
   \times\Bigg\{ &
\frac{1}{(z_0+z_2)^2} \left[ 2z_0(z_0+z_2)+z_2^2 \right] \left[ 1-2 z_1 (1-z_1) \right] \frac{|\xt_{3;(\textrm{j})}|^2 |\xt_{2;(\textrm{j})}|^2}{256}
\overline g_{(\textrm{j})}^{(2;2)}( y_1)  \overline g_{(\textrm{j})}^{(2;2)}( y_2)   \\
+&\frac{1}{(z_1+z_2)^2} \left[ 2z_1(z_1+z_2)+z_2^2 \right] \left[1-2z_0(1-z_0) \right]\frac{|\xt_{3;(\textrm{k})}|^2 |\xt_{2;(\textrm{k})}|^2}{256}
\overline g_{(\textrm{k})}^{(2;2)}( y_1)  \overline g_{(\textrm{k})}^{(2;2)}( y_2) \Bigg\}
    \end{split}
\end{equation}

\begin{equation}
\label{eq:qqg_T_3m}
    \begin{split}
        I^T_{3m}  =&  \int_0^1 \dd{y_{1}} \int_0^1 \dd{y_{2}}  m^2 \frac{1}{z_2}  \\
      \times  \Bigg\{&
\frac{1}{(z_0+z_2)^2} \left[ 2z_0(z_0+z_2)+z_2^2 \right] 
\frac{ |\xt_{2;(\textrm{j})}|^2}{64}
\overline g_{(\textrm{j})}^{(1;2)}( y_1)  \overline g_{(\textrm{j})}^{(1;2)}( y_2) \\
+&\frac{1}{(z_1+z_2)^2} \left[ 2z_1(z_1+z_2)+z_2^2 \right] 
\frac{ |\xt_{2;(\textrm{k})}|^2}{64}
\overline g_{(\textrm{k})}^{(1;2)}( y_1)  \overline g_{(\textrm{k})}^{(1;2)}( y_2) 
\Bigg\}
    \end{split}
\end{equation}

\begin{equation}
\label{eq:qqg_T_3F}
    \begin{split}
&I^T_{3F} = 
 \int_0^1 \dd{y_{1}} \int_0^1 \dd{y_{2}} 
         \frac{1}{2} \frac{1}{z_2}
\overline g_{(\textrm{j})}^{(2;2)}(y_1)  \overline g_{(\textrm{k})}^{(2;2)}(y_2) \\
 &\times 
\frac{1}{64(z_0+z_2)(z_1+z_2)} 
\left\{ 
z_2 (z_0-z_1)^2 \left[ \left( \xt_{3;(\textrm{j})} \vdot \xt_{2;(\textrm{j})} \right)\left( \xt_{3;(\textrm{k})} \vdot \xt_{2;(\textrm{k})} \right)-\left( \xt_{3;(\textrm{j})} \vdot \xt_{2;(\textrm{k})} \right)\left( \xt_{3;(\textrm{k})} \vdot \xt_{2;(\textrm{j})} \right) \right] \right.\\
&\left.-\left[z_1(z_0+z_2)+z_0(z_1+z_2) \right] \left[z_0(z_0+z_2)+z_1(z_1+z_2) \right] \left( \xt_{2;(\textrm{j})} \vdot \xt_{2;(\textrm{k})} \right)\left( \xt_{3;(\textrm{j})} \vdot \xt_{3;(\textrm{k})} \right)
\right\} 
    \end{split}
\end{equation}

\begin{equation}
\label{eq:qqg_T_3Fm}
    \begin{split}
&I^T_{3F_m}  = 
\int_0^1 \dd{y_1} \int_0^1 \dd{y_2}  \frac{1}{z_2} \frac{m^2}{2} \\
&\times \Bigg\{
\frac{z_2^4}{64 (z_0+z_2)^4} \left[4 z_1(z_1-1)+2 \right] |\xt_{3;(\textrm{j})}|^2
g_{(\textrm{j})}^{(2;1)}(y_1) 
g_{(\textrm{j})}^{(2;1)}(y_2) \\
&-\frac{z_0z_1z_2^2}{16(z_0+z_2)^3} \left( \xt_{3;(\textrm{j})} \vdot \xt_{2;(\textrm{j})} \right)
\overline g_{(\textrm{j})}^{(1;2)}(y_1) 
g_{(\textrm{j})}^{(2;1)}(y_2)\\
&+\frac{z_2^4}{64 (z_1+z_2)^4} \left[4 z_0(z_0-1)+2 \right] |\xt_{3;(\textrm{k})}|^2 
g_{(\textrm{k})}^{(2;1)}(y_1) 
g_{(\textrm{k})}^{(2;1)}(y_2) \\
&+\frac{z_0z_1z_2^2}{16(z_1+z_2)^3} \left( \xt_{3;(\textrm{k})} \vdot \xt_{2;(\textrm{k})} \right)
\overline g_{(\textrm{k})}^{(1;2)}(y_1) 
g_{(\textrm{k})}^{(2;1)}(y_2)\\
&- \frac{1}{32(z_0+z_2)(z_1+z_2)} \left[(2z_0+z_2)(2z_1+z_2)+z_2^2 \right] \left( \xt_{2;(\textrm{j})} \vdot \xt_{2;(\textrm{k})} \right)
\overline g_{(\textrm{j})}^{(1;2)}(y_1) 
\overline g_{(\textrm{k})}^{(1;2)}(y_2)\\
&+ \frac{z_2^4}{32(z_0+z_2)^2(z_1+z_2)^2} \left[(2z_0+z_2)(2z_1+z_2)+z_2^2 \right] \left( \xt_{3;(\textrm{j})} \vdot \xt_{3;(\textrm{k})} \right)
g_{(\textrm{j})}^{(2;1)}(y_1) 
g_{(\textrm{k})}^{(2;1)}(y_2) \\
&+\frac{m^2}{16} \frac{2 z_2^4}{(z_0+z_2)^4}
g_{(\textrm{j})}^{(1;1)}(y_1) 
g_{(\textrm{j})}^{(1;1)}(y_2)
+\frac{m^2}{16} \frac{2 z_2^4}{(z_1+z_2)^4}
g_{(\textrm{k})}^{(1;1)}(y_1) 
g_{(\textrm{k})}^{(1;1)}(y_2) \\
& -\frac{(z_0z_2)^2}{16(z_0+z_2)(z_1+z_2)^2} \left( \xt_{2;(\textrm{j})} \vdot \xt_{3;(\textrm{k})} \right)
\overline g_{(\textrm{j})}^{(1;2)}(y_1) 
g_{(\textrm{k})}^{(2;1)}(y_2)\\
&+\frac{(z_1z_2)^2}{16(z_0+z_2)^2(z_1+z_2)} \left( \xt_{3;(\textrm{j})} \vdot \xt_{2;(\textrm{k})} \right)
\overline g_{(\textrm{k})}^{(1;2)}(y_1) 
g_{(\textrm{j})}^{(2;1)}(y_2) \\
& -\frac{z_0z_1 z_2^2}{16(z_0+z_2)^3}  \left( \xt_{2;(\textrm{j})} \vdot \xt_{3;(\textrm{j})} \right) 
g_{(\textrm{j})}^{(1;1)}(y_1) 
\overline g_{(\textrm{j})}^{(2;2)}(y_2) 
+\frac{z_0z_1 z_2^2}{16(z_1+z_2)^3}  \left( \xt_{2;(\textrm{k})} \vdot \xt_{3;(\textrm{k})} \right) 
g_{(\textrm{k})}^{(1;1)}(y_1) 
\overline g_{(\textrm{k})}^{(2;2)}(y_2)\\
& - \frac{(z_0+z_2) z_2^2}{16(z_1+z_2)^2}   \left( \xt_{2;(\textrm{j})} \vdot \xt_{3;(\textrm{j})} \right)
g_{(\textrm{k})}^{(1;1)}(y_1) 
\overline g_{(\textrm{j})}^{(2;2)}(y_2)
+\frac{(z_1+z_2) z_2^2}{16(z_0+z_2)^2}   \left( \xt_{2;(\textrm{k})} \vdot \xt_{3;(\textrm{k})} \right)
g_{(\textrm{j})}^{(1;1)}(y_1) 
\overline g_{(\textrm{k})}^{(2;2)}(y_2)
\Bigg\}. 
    \end{split}
\end{equation}
The sum of these  contributions with one additional numerical integration is evaluated in
\begin{cppcode}
double ITNLOqg_massive_tripole_part_I2(double Q2, double mf, double z1, double z2, double x01sq, double x02sq, double x21sq, double y_t)
\end{cppcode}
and the sum of contributions with two additional numerical integrals in
\begin{cppcode}
double ITNLOqg_massive_tripole_part_I3(double Q2, double mf, double z1, double z2, double x01sq, double x02sq, double x21sq, double y_t1, double y_t2) 
\end{cppcode}

\end{appendix}





\bibliography{SciPost_Example_BiBTeX_File.bib}


\end{document}